\documentclass[
preprint,
superscriptaddress,
]{revtex4-2}

\usepackage{gensymb}
\usepackage{graphicx}   
\usepackage{dcolumn}   

\usepackage{bm}         
\usepackage{color}
\usepackage{multirow}
\usepackage{placeins} 

\usepackage{amsmath}
\usepackage{amssymb}
\begin{document}

\title{Molecular origin of slippery behaviour in tethered liquid layers}

\author{Fabio Rasera}
\thanks{These authors contributed equally}
\affiliation{Dipartimento di Ingegneria Meccanica e Aerospaziale, Sapienza Universit\`a di Roma, 00184, Rome, Italy}
\author{Isaac J. Gresham}
\thanks{These authors contributed equally}
\affiliation{School of Chemistry, The University of Sydney, NSW 2006, Australia}
\affiliation{The University of Sydney Nano Institute, The University of Sydney, NSW 2006 Australia}
\author{Antonio Tinti}
\affiliation{Dipartimento di Ingegneria Meccanica e Aerospaziale, Sapienza Universit\`a di Roma, 00184, Rome, Italy}
\author{Chiara Neto}
\email{chiara.neto@sydney.edu.au}
\affiliation{School of Chemistry, The University of Sydney, NSW 2006, Australia}
\affiliation{The University of Sydney Nano Institute, The University of Sydney, NSW 2006 Australia}
\author{Alberto Giacomello}
\email{alberto.giacomello@uniroma1.it}
\affiliation{Dipartimento di Ingegneria Meccanica e Aerospaziale, Sapienza Universit\`a di Roma, 00184, Rome, Italy}

\date{\today}

\begin{abstract}
 Slippery covalently attached liquid surfaces (SCALS) are a family of nanothin polymer layers with remarkably low static droplet friction, characterised by a low contact angle hysteresis (CAH$< 5 \degree$), which makes them ideally suited to self-cleaning, water harvesting, and anti-fouling applications.
 Recently, a Goldilocks zone of lowest CAH has been identified for polydimethyl siloxane (PDMS) SCALS of intermediate thickness (3.5 nm), yet, molecular-level insights are missing to reveal the underlying physical mechanism of this elusive, slippery optimum.
 In this work, the agreement between coarse-grained molecular dynamics simulations and atomic force microscopy data shows that nanoscale defects, as well as deformation for thicker layers, are key to explaining the existence of this `just right' regime. At low thickness values, insufficient substrate coverage gives rise to chemical patchiness; at large thickness values, two features appear: 1) a waviness forms on the surface of the liquid layer due to a previously overlooked lateral microphase separation occurring in polydisperse brushes, and 2) layer deformation due to the contact line is larger than in thinner layers. The most pronounced slippery behaviour occurs for smooth PDMS layers that do not exhibit nanoscale waviness. The converging insights from molecular simulations, experiments, and a contact angle hysteresis theory provide design guidelines for tethered polymer layers with ultra-low contact angle hysteresis. 

\end{abstract}


\maketitle

\newpage
\section{\label{sec:intro}Introduction}

    Slippery Covalently Attached Liquid Surfaces (SCALS) are an emerging class of materials consisting of chemically attached liquid polymer chains (i.e. glass transition temperature $T_\mathrm{g}$ below ambient) \cite{gresham2023advances, chen2023omniphobic}.
    SCALS exhibit exceptional properties, such as easy droplet shedding, self-cleaning, ice-shedding, and anti-biofouling, often equivalent to lubricant-infused \cite{Peppou2020} and superhydrophobic surfaces \cite{Scarratt2017}. 
    SCALS are regarded as more robust than other super-wettability surfaces \cite{giacomello2016perpetual}, as their performance does not rely on a specific surface texture or an easily depleted infused liquid lubricant \cite{chen2023omniphobic}.
    To date, the most promising SCALS are made of polydimethylsiloxane (PDMS), which is non-toxic, FDA approved, and harmlessly degrades in environmental conditions, and therefore is applicable in a wide range of contexts, from medical devices to atmospheric water capture.
    SCALS are timely eco-friendly coatings that could replace some perfluorinated `forever chemicals' in anti-fouling \cite{Chen2021PDMSBetterAntiScale, Zhao2020NonFluoroPaper, Shabanian2024SustainableDesign, Hauer2024WettingSilicone}. 

    The exceptional performance of SCALS is quantified by measuring contact angle hysteresis (CAH, the difference between advancing and receding contact angle), which is proportional to droplet static friction \cite{mchale2022} and correlates well with desirable properties such as anti-fouling and anti-scaling \cite{gresham2023advances}.
    The extremely low CAH ($\approx 1\degree$), and the corresponding high performance possible with SCALS, superior to alkyl self-assembled monolayers \cite{Lepikko2024}, is attributed to their `liquid like' nature. 
    
\begin{figure}[ht]
        \centering
        \includegraphics[width=.85\linewidth]{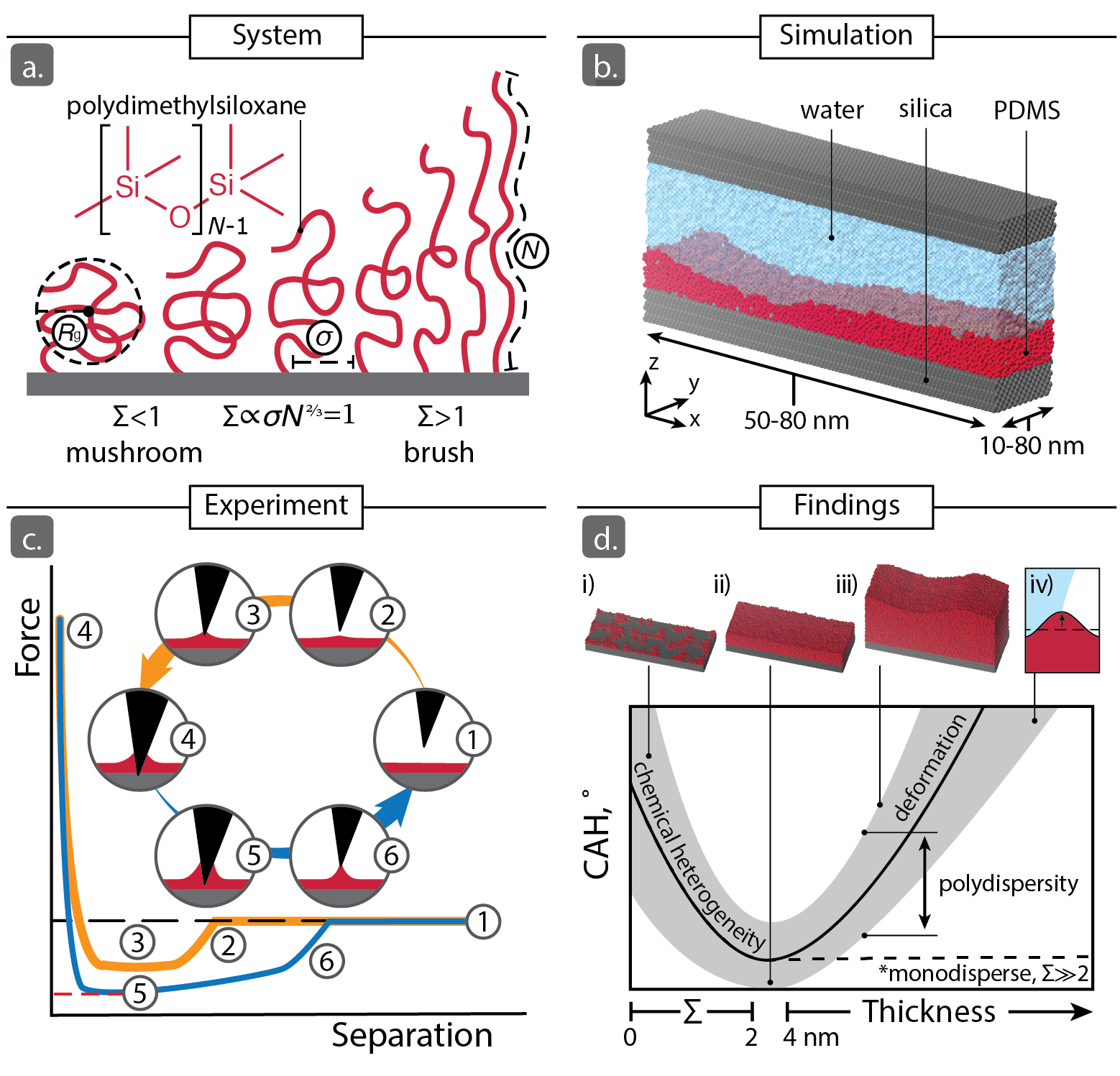}
        \caption{\textbf{Combined simulation and experimental study of CALS}
        a) CALS consist of end-tethered PDMS chains, which can be defined by their grafting density $\sigma$ and average chain length $\overline{N}$. Reduced grafting density, $\Sigma$, is a powerful predictor of slippery performance.
        b) Example of a simulated system, which predicts that long chain, polydisperse CALS form wavy structures.
        c) Meniscus force mapping can distinguish the liquid layer from the underlying solid surface.
        d) Schematic illustrating the literature trend and identifying the mechanisms responsible: i) CAH of thin layers is dominated by chemically heterogeneity, best described by $\Sigma$; ii) the lowest CAH is obtained for smooth layers without waviness and deformation; iii)-iv) CAH of thick layers is dominated by layer deformation, which tends to scale with thickness.
        At all thickness values, polydispersity increases CAH by increasing layer waviness (iii).
        Monodisperse, high density layers are expected to have low CAH at all thickness values, but they have not been produced experimentally.
        \label{fig:scals}}
\end{figure}

    SCALS belong to the broader family of end-tethered polymer layers, which, at high grafting density, form  polymer brushes.
    The defining feature of SCALS is that the tethered polymers are in a liquid state and hence do not require solvation to be effective.
    This is in contrast to traditional polymer brushes; for example, polyethylene oxide brushes are only expected to be antifouling when hydrated.
    As shown schematically in Fig.~\ref{fig:scals}a, the structure and behaviour of tethered polymers is often explained by a single parameter: the reduced grafting density, $\Sigma$, which is a measure of the degree of chain crowding \cite{Brittain2007StructuralDefinitionPolymer}:
    \begin{equation} \label{eq:Sigma}
        \Sigma = \sigma \pi R_\mathrm{g}^2 \, ,
    \end{equation}
    where $R_\mathrm{g}$ is the average radius of gyration of the grafted polymers (dependent on molecular weight and solvent quality) and $\sigma$ is the grafting density (grafting points per unit area, $\mathrm{gps}/\mathrm{nm}^{2}$). $\Sigma < 1$ corresponds to a sparse (pancake/mushroom) regime, while $\Sigma \gg 1$ corresponds to a dense (brush) regime.

\FloatBarrier
    
   Extensive literature has identified a `Goldilocks zone' of lowest CAH at intermediate layer thickness  and/or molecular weight, as schematically shown by the parabolic trend in Fig.~\ref{fig:scals}d. \cite{Wang2016SOCAL, Krumpfer2011RediscoveringSiliconesUnreactive, Flagg2017, Xiaoteng2024, Sarma2022, Lhermerout2019, chen2023omniphobic,gresham2023nanostructure}.  The shaded region around the parabola shows that, both within and between studies, apparently similar PDMS layers (in thickness, composition, contact angle) have different CAH values; extremes between $1\degree$ (corresponding to slippery surfaces) and $15\degree$  (non-slippery, hence referred to as CALS) have been reported \cite{gresham2023advances,Xiaoteng2024,chen2023omniphobic}. 
The molecular origins of these drastic performance differences are only beginning to be understood. Recently Gresham \textit{et al.} demonstrated that $\Sigma$ is a good predictor for the Goldilocks region for SCALS prepared with different methods, with minimum CAH obtained around $\Sigma=2$ \cite{gresham2023nanostructure}.
    From this summary of the state of the art, two outstanding questions emerge:
    \begin{enumerate}
        \item What causes the widely-observed parabolic trend and corresponding Goldilocks zone?
        \item Why do otherwise identical PDMS layers exhibit different CAH?
    \end{enumerate}

    In this work coarse-grained molecular dynamics (CG-MD) simulations, informed by experimentally measured physicochemical parameters (chain length, polydispersity, grafting density) \cite{gresham2023nanostructure}, were used to address these two questions.
    Key findings from the simulations are shown in Fig.~\ref{fig:scals}d: i) low $\Sigma$ CALS are chemically heterogeneous, with patches of the substrate exposed, while iii) in thicker layers, polydispersity induces nanoscale lateral waviness.
    Low CAH surfaces are ii) chemically uniform and smooth.
    These simulation features were experimentally observed by adapting atomic force microscopy meniscus force mapping (MFM) to these nano-thin PDMS layers (Fig.~\ref{fig:scals}c). 
    Simulations found that thick layers deform at the droplet three-phase contact line (iv).
    This deformation has recently been identified as a potential cause of hysteresis in such layers \cite{Xiaoteng2024}, with comparisons made to wetting on macroscopically soft surfaces (i.e., cross-linked PDMS gels) \cite{Hauer2024WettingSilicone}.
    By applying simple wetting models to the simulated structures, mechanistic explanations for the questions posed above are offered.

\section{\label{sec:results}Results}

   The key strength of this work is the combination and close agreement between simulations and experiments, with key layer properties shown in Table~\ref{tab:Params}. Using CG-MD, grafted PDMS layers characterised by different chain length ($N$), chain length distribution (polydispersity, PDI) and grafting density ($\sigma$) were simulated in vacuum and in water, which are both poor solvents for PDMS. A CG force field specific for PDMS was used, which allowed modelling of realistic PDMS behaviour \cite{cambiaso2022development}. PDMS molecules were grafted to a flat, silica-like substrate \cite{cambiaso2023grafting}, and allowed to equilibrate from the fully elongated configuration. Different realisations of chain lengths and grafting points were simulated. The chain length distribution in simulations mimicked that measured in experiments (see Fig.~S1). Simulated layers named P1, P2, and P3 in Table~\ref{tab:Params} correspond to low, intermediate and high values of average polymer repeat units ($\overline{N}$), respectively, and are polydisperse (PDI = 1.3-1.35).
   For all $\overline{N}$, a range of experimentally relevant grafting densities were simulated (Table~\ref{tab:Params}).
   System P2, of intermediate thickness and chain length, corresponds to the SCALS with the lowest CAH ($2.7\degree$) as previously reported \cite{gresham2023nanostructure}, while systems P1 and P3 are not slippery in experiments. Three monodisperse layers, M50, M100, and M300 (PDMS chains with 50, 100, and 300 monomers) were also simulated for comparison, as shown in the Supplementary information and in Fig.~S2 and Fig.~S3.

\begin{table}[]
\caption{Structural parameters of grafted PDMS layers studied in this work in both simulations and experiments: $\overline{N}$ = average number of repeat units in the polymer chains, PDI = chain polydispersity, $\sigma$ = grafting density and $d$ =layer thickness. Simulated layers P1 - P3 are informed by experimental parameters and are polydisperse, while layers M50 - M300 are idealised and monodisperse. Sim. small and Sim. big refer to simulation base areas of 50x10 and 80x80 nm$^2$, respectively.}

\label{tab:Params}
\scalebox{0.8}{
\begin{tabular}{cccccc}
\multicolumn{2}{c}{Sample}                                & $\overline{N}$                  & PDI                   & $\sigma$, gps/nm$^2$ & $d$, nm                              \\ \hline
\multirow{3}{*}{P1} & \multicolumn{1}{c|}{Sim. small}     & \multirow{2}{*}{15}  & \multirow{2}{*}{1.3}  & 0.1/0.2/0.3/0.6/1.0         & 0.8$\pm$1/0.9$\pm$0.2/1.0$\pm$0.2/1.3$\pm$0.2/2.1$\pm$0.2   \\
                    & \multicolumn{1}{c|}{Sim. big}       &                      &                       & 0.3                  & 0.9$\pm$0.2                          \\
                    & \multicolumn{1}{c|}{Exp.$^\dagger$} & 8/NA                 & 1.4/NA                & 0.9/NA               & 0.9/1.1                              \\ \hline
\multirow{3}{*}{P2} & \multicolumn{1}{c|}{Sim. small}     & \multirow{2}{*}{88}  & \multirow{2}{*}{1.3}  & 0.1/0.2/0.3/0.6/1.0          & 1.9$\pm$0.4/2.6$\pm$0.4/3.7$\pm$0.5/6.8$\pm$0.2/11.7$\pm$0.3  \\
                    & \multicolumn{1}{c|}{Sim. big}       &                      &                       & 0.3                  & 3.5$\pm$0.5                          \\
                    & \multicolumn{1}{c|}{Exp. (3 nm) $^\dagger$} & 32/88                & 1.1/1.3               & 0.75/0.26            & 3.1/2.9                              \\ 
                    & \multicolumn{1}{c|}{Exp. (5 nm) $^\dagger$} & 91/-                 & 1.1/-                 & 0.51/-               & 5.94/-                              \\ \hline
\multirow{3}{*}{P3} & \multicolumn{1}{c|}{Sim. small}     & \multirow{2}{*}{308} & \multirow{2}{*}{1.35} & 0.1/0.2/0.3         & 4$\pm$1/8$\pm$1/12$\pm$2            \\
                    & \multicolumn{1}{c|}{Sim. big}       &                      &                       & 0.3                  & 12.0$\pm$0.8                         \\
                    & \multicolumn{1}{c|}{Exp.$^\dagger$} & 142/308              & 2.3/1.4               & 0.44/0.21            & 8.1/8.1                              \\ \hline
M50                 & \multicolumn{1}{c|}{Sim. small}     & 50                   & 1                     & 0.1/0.2/0.3/0.6/1.0          & 1.4$\pm$0.5/1.8$\pm$0.5/2.2$\pm$0.7/4.0$\pm$0.5/6.5$\pm$0.3  \\
M100                & \multicolumn{1}{c|}{Sim. small}     & 100                  & 1                     & 0.1/0.2/0.3/0.6/1.0          & 2.0$\pm$0.8/3$\pm$1/4.0$\pm$0.2/7.8$\pm$0.3/12.9$\pm$0.4      \\
M300                & \multicolumn{1}{c|}{Sim. small}     & 300                  & 1                     & 0.1/0.2/0.3          & 4.0$\pm$0.5/7.9$\pm$0.3/11.6$\pm$0.3
\end{tabular}

}

\footnotesize{$^\dagger$ For experimental systems, two values are reported, respectively: measured by (reflectometry + numerical self-consistent field theory)/(measured by ellipsometry + single-molecule force spectroscopy). Experimental values shown here are obtained from \cite{gresham2023nanostructure}.}
\end{table}

\begin{figure}[ht]
        \centering
        \includegraphics[width=1\linewidth]{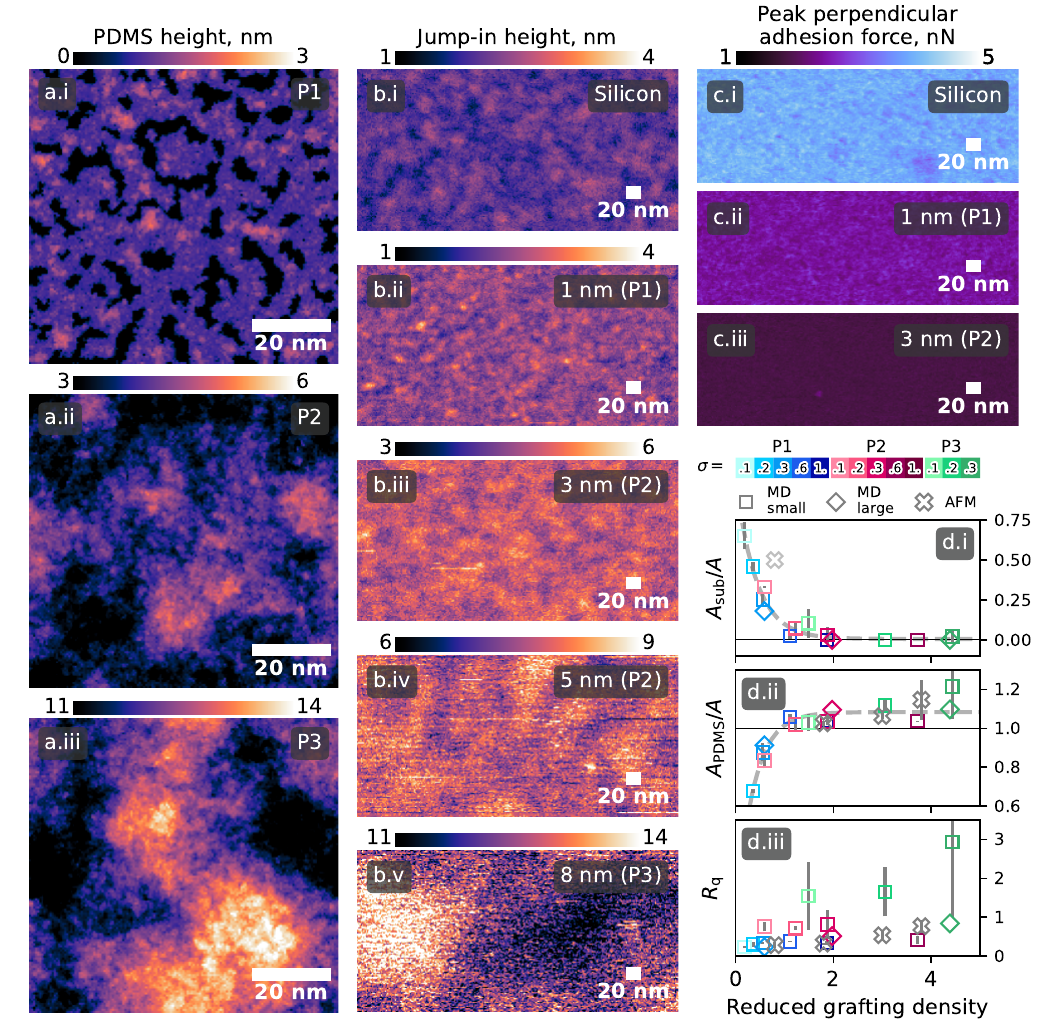}
        \caption{  \textbf{Surface topography of PDMS CALS}:
         (a) predicted by MD and (b) measured by MFM, for increasing $\Sigma$ (Table~\ref{tab:Params}).
        (b.i) MFM image of unmodified silicon wafer (non-zero jump-in due to humidity).
        (c) Peak adhesion maps from MFM. P1, the low $\Sigma$  PDMS layer, has a chemically heterogeneous surface, with peak adhesion values between that of silica (c.i) and that of thicker PDMS layers (c.iii).
        (d) Parameters derived from panels a-c, plotted as a function of $\Sigma$: (d.i) area of silica accessible to the contacting liquid divided by the projected surface area, (d.ii) area of PDMS accessible to the contacting liquid divided by the projected surface area, and (d.iii) RMS roughness of the PDMS layers. Colour corresponds to $N$ and $\sigma$; markers denote values derived from $50 \times 10$~nm$^2$ MD boxes ($\square$), $80 \times 80$~nm$^2$ MD boxes ($\diamond$), and AFM images ($\times$).}
        \label{fig:dens_profiles}
\end{figure}
    \FloatBarrier
    
    The experimental PDMS layers were prepared using the synthetic method of Krumpfer and McCarthy \cite{Krumpfer2011RediscoveringSiliconesUnreactive}. Highly detailed data on layer structure was obtained from previous work by Gresham \textit{et al.} \cite{gresham2023nanostructure}, as shown in Table~\ref{tab:Params}.
    CAH was assessed by adding/withdrawing volume from a water droplet quasi-statically and measuring the advancing/receding contact angle.
    New experimental data consists of AFM maps of topography and adhesion force for the grafted PDMS layers. The Bruker Multimode Peakforce Imaging AFM mode was used to produce force–distance curves at every point of the scanned area.
    From the force-distance curves, meniscus force measurements (MFM) were taken: the position of the jump-in at which the liquid meniscus touches the AFM tip was extracted using a custom script, and this allowed to reconstruct the nanoscale topography of the liquid layer \cite{Peppou2018}. A similar approach has been developed by Zhou et al. \cite{Xiaoteng2024}.
    Peakforce imaging was used as it was challenging to accurately track the surface of these liquid layers with conventional tapping-mode AFM, as discussed further in Supplementary Information.
    
\subsection{Chemical and topographical nanoscale defects in grafted surfaces}

The CALS topography produced by MD simulations and measured by MFM are presented in Fig.~\ref{fig:dens_profiles}, and show experiment and simulation in excellent agreement.
Broadly, both simulation and experiments featured chemically heterogeneous layers at low $\Sigma$ (Fig.~\ref{fig:dens_profiles}a.i, b.ii), which transitioned into chemically homogeneous layers at intermediate $\Sigma$ (Fig.~\ref{fig:dens_profiles}a.ii, b.iii, b.iv).
As $\Sigma$ increased further, the surfaces remained chemically homogeneous, but an unexpected topographical waviness of low aspect ratio emerged in both simulations and experiments (Fig.~\ref{fig:dens_profiles}a.iii, b.v). Simulations show that such waviness is always in full contact with water.
The trends observed in the presented images are supported by additional simulations, summarised in Fig.~\ref{fig:dens_profiles}d.
Figure~\ref{fig:dens_profiles}d.i shows that at $\Sigma < 2$ the area of the substrate accessible to the contacting liquid ($A_\mathrm{sub}$) is a significant fraction of the entire projected surface area ($A$), indicative of a chemically heterogeneous surface.
At higher $\Sigma$ values, Fig.~\ref{fig:dens_profiles}d.ii demonstrates that the surface area of the PDMS layer ($A_\mathrm{PDMS}$) exceeded that of the planar substrate, corresponding to the appearance of waviness.
The formation and growth of these wave-like features is also captured by the increase in RMS roughness ($R_\mathrm{q}$) with $\Sigma$ observed in Fig.~\ref{fig:dens_profiles}d.iii.
Results derived from simulations (colored markers) match those measured experimentally (grey crosses), both in absolute terms and in the trends observed.

A detailed comparison of simulated and experimental results in Fig.~\ref{fig:dens_profiles} is provided here, starting at low $\Sigma$ and proceeding to higher values. The simulated low chain length surface (P1) was patchy at $\sigma$ values below $0.6$~gps/nm$^2$, with large fractions of the bare substrate (between $70$ and $15\%$, shown in black in Fig.~\ref{fig:dens_profiles}a.i) exposed to the contacting liquid  (blue data points in Fig.~\ref{fig:dens_profiles}d).
As expected, the exposed substrate area decreased with increasing grafting density, with full coverage only reached at the highest $\sigma = 1~ \mathrm{gps}/\mathrm{nm}^{2}$. Overall, the P1 surface showed strong chemical heterogeneity in simulations, but the patches of bare silica could not be resolved by AFM due to their small size.
However, adhesion of the AFM tip to the surface could be used as a proxy of exposed substrate; Fig.~\ref{fig:dens_profiles}c shows that the thin PDMS layer in P1 (c.ii) had intermediate adhesion ($\approx3.2$~nN) between that of an uncoated silica native oxide layer (c.i, $\approx4.5$~nN) and that of a thicker PDMS layer (c.iii, $\approx2$~nN).
From these measurements, a coverage of 50\% is approximated and plotted in Fig.~\ref{fig:dens_profiles}d.i as a grey cross.
Maps of the peak adhesion value for P1 (Fig.~\ref{fig:dens_profiles}c.ii) were more heterogeneous than those for thicker PDMS layers (Fig.~\ref{fig:dens_profiles}c.iii), supporting the presence of the chemical defects seen in the computation. 

The simulated intermediate chain length surface (P2) showed lower chemical heterogeneity than P1, with a significant fraction of exposed substrate observed  only at the lowest grafting density,  $\sigma=0.1~\mathrm{nm}^{-2}$ (Fig.~\ref{fig:dens_profiles}d.i).
Surprisingly, the layer topography was not entirely smooth even for full substrate coverage, as might have been expected for a liquid interface.
Instead, the emergence of waviness was observed, with an amplitude on the scale of 1~nm and a wavelength of approximately 20~nm (Fig.~\ref{fig:dens_profiles}a.ii). 
These features became more pronounced as layer thickness increased further. For the high chain length surface (P3) in Fig.~\ref{fig:dens_profiles}a.iii, the waviness had an amplitude of $\approx5$~nm and wavelength around $50$~nm, and was observed at all grafting density values studied, leading to an increase of $A_\mathrm{PDMS}/A$  above $1$ and to a growing $R_q$ (Fig.~\ref{fig:dens_profiles}d.ii and d.iii).
AFM maps of comparable surfaces confirmed this surface topography (Fig.~\ref{fig:dens_profiles}b.iii), although, for layers with comparable thicknesses and/or $\Sigma$, the wavy features observed experimentally had larger wavelength and slightly lower amplitude than in simulations.
Still, the overall trend remained remarkably close, with waviness size increasing as $\Sigma$ increased.

\clearpage

The waviness observed here is a static feature, both in simulations and experiments, and as such cannot be rationalised as capillary waves, as have been observed at liquid-air interfaces.
Further investigation of the origin of these features (discussed below) confirm they are expected to be static.
Furthermore, the MFM technique enables the simultaneous determination of the substrate and layer topography, which allowed us to exclude that substrate topography is at the origin of the wave-like features observed.
Furthermore, maps of the maximum adhesion force became uniform as $\Sigma$ increased   (Fig.~\ref{fig:dens_profiles}c.iii), indicating that the higher $\Sigma$ layers were chemically homogeneous.

The waviness predicted by MD simulations and revealed by MFM is the most unexpected element of Fig.~\ref{fig:dens_profiles}.
Because the polymers are above their $T_\mathrm{g}$ ($T_\mathrm{g}$$<-100^\circ$C for PDMS) and grafted relatively sparsely, it has been argued that the layer would self-smooth on all length scales,\cite{Lhermerout2016} as would be observed for a true liquid interface.
However, these tethered-liquid polymer layers are shown to be heterogeneous both in terms of their chemistry (at  $\Sigma<2$) and of their topography ($\Sigma>2$). 
From this observation two questions arise: what is the origin of these features, and can they explain the Goldilocks minimum in CAH observed across the literature?
Both questions are discussed in turn in the following sections.

\subsection{\label{sec:uphase_separation} Polydispersity induces waviness}
The nanoscale waviness observed for polydisperse cases is reminiscent of the phenomenology of microphase separation, common in block copolymers and mixed polymer brushes \cite{lee2012interfacial,minko2002lateral,muller2002phase}.
However, to the best of our knowledge, this microphase separation has not been previously reported for liquid-like grafted polymers of a single monomer type.
Below, these features are shown to be due to polymer segregation based on chain length, which occurs both vertically and laterally.
While these segregation modes are inexorably linked, vertical separation will be discussed first, as it has previously been predicted by theoretical approaches, before turning to the more novel lateral separation of homopolymers.

    \begin{figure}[t]
        \centering
        \includegraphics[width=0.5\linewidth]{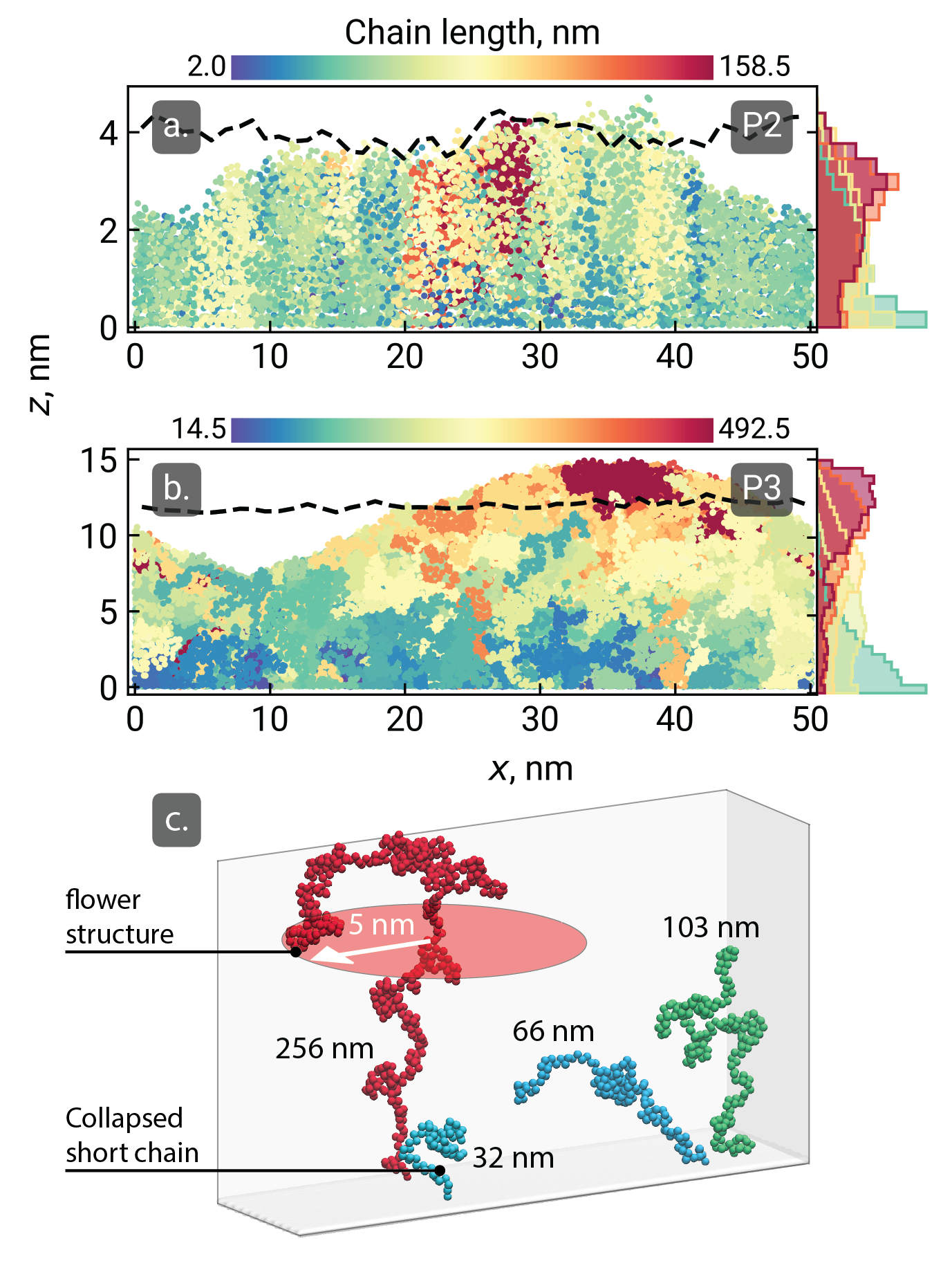}
        \caption{\textbf{Waviness is caused by length-wise segregation of chains:}
        a,b) cross-sectional profiles of P2 and P3 layers ($\sigma=0.3~\mathrm{nm}^{-2}$), with chains colored according to contour length.
        Different chain length fractions segregate to form wavy features;
        vertical chain segregation is captured by the accompanying histogram, which compares favorably to theoretical results (Fig.~S4) \cite{devos2009modeling}.
        Comparable monodisperse layers (M100 and M300, respectively) do not develop waves; their surfaces are plotted as dashed lines.
        c) Conformations of different length chains in b, showing the flower structure of long chains (red).
        The $R_g$ of chains in the flower regime (red circle) is not large enough to cover the whole surface, hence the emergence of waviness. }
        \label{fig:chain_lengths}
    \end{figure}

Vertical phase separation can be clearly seen in the histograms adorning the side of Fig.~\ref{fig:chain_lengths}a and b, which shows the segregation of chains based on their contour length.
Short chains coil close to the substrate (even though their contour length is sufficient for them to extend throughout the layer; compare colorbar and y-axes scale), intermediate chains are relatively stretched and dispersed throughout the layer, and longer chains stretch to the top of the layer (Fig.~\ref{fig:chain_lengths}c), where their apical segments coil in flower-like aggregates.
This partitioning of longer chains to the brush-solvent interface (named flower-stem) is predicted by numerical self-consistent field theory (nSCFT) \cite{devos2009modeling, milner1988, milner1989effects}, a mean-field technique that successfully described polymer brush behaviour in a number of contexts \cite{gresham2023nanostructure, Johnson2020nSCFTBrush,Gresham2021GeometricalConfinementModulates, Abbott2015IsOsmoticPressure}.
de Vos and Leermakers \cite{devos2009modeling} model the structure of a polydisperse brush with nSCFT and plot the volume fractions of polymers with different lengths, which nicely match the histograms in Fig.~\ref{fig:chain_lengths}a and b (see Fig.~S4); fundamental work on bi-disperse brushes also report similar phenomena \cite{lai1992,lai1995,klushin1991}.
The excellent agreement between the vertical phase separation predicted by our simulation and nSCFT support our subsequent observation of lateral separation. 

Figure~\ref{fig:chain_lengths}b shows that the vertical and lateral segregation are linked: longer chains concentrate at the bulge of the waves, medium chains at the troughs, and shorter chains at the bottom of the layer.
This behaviour can be rationalised by considering a bidisperse brush, in which the longer polymers extend past the shorter polymers (the stem) and spread into a flat (the crown) aggregate.
In the same way, in polydisperse systems, chains sufficiently longer than the average coil over the rest of the layer as the effective chain crowding decreases far from the substrate.
The characteristic size of the bumps is determined by the gyration radius of the `flower' segments of the longer chains $R_{g,f}$ above the average brush thickness \cite{lai1995} (Fig.~\ref{fig:chain_lengths}c). If the number of flower structures is not sufficient to entirely cover the surface, i.e., if the reduced grafting density of flowers $N_f \times R_{g,f}^2/A<1$, then surface waviness must emerge. 

Given the mechanism behind waviness, both the amplitude and the wavelength of the features in Fig.~\ref{fig:dens_profiles} are expected to depend on $R_{g,f}$ and thus to increase with increasing mean chain length. 
For instance, in P3, chains have lengths ranging up to ca. $500$~nm with an average of $300$~nm, while for P2 the maximum is only $125$~nm, with an average of $90$~nm. Additionally, going from P1 to P3 a progressive stretching of the PDMS layer is observed in the middle of the layer, while the top of the layer is coiled (Fig.~\ref{fig:chain_lengths}c).
Monodisperse brushes do not develop the waviness predicted for their polydisperse counterparts (smooth surfaces plotted in Fig.~\ref{fig:chain_lengths}a, b as dashed lines).

In summary, Fig.~\ref{fig:chain_lengths} shows that the emergence of waviness in sufficiently thick, polydisperse PDMS layers is related to the way in which chains of different length self-organise in the layer.

\subsection{\label{sec:CAH} Making sense of contact angle hysteresis}

It is well established that both chemical defects (as in P1, Fig.~\ref{fig:dens_profiles}) and topographical roughness (the waviness most pronounced in P3, Fig.~\ref{fig:dens_profiles}) result in CAH \cite{Joanny1984,butt2022}.
As the P2 sample, which corresponds to the lowest CAH experimental sample, appears to have no chemical defects (compared to P1) and less waviness (compared to P3), it is tempting to use these features to answer the two questions posed in the introduction.
To do so quantitatively, a simple theory linking chemical defects and waviness to CAH is presented below.

The classical model for static CAH by Joanny and de~Gennes \cite{Joanny1984} and its extensions relate CAH to the number of defects and to the energy dissipated by a defect in an advancing/receding cycle \cite{Joanny1984,Crassous1994,Ramos2003}. 
The nanoscale defects present in CALS can collectively give rise to hysteresis, because each of them may exert a small pulling (upon receding) or pushing force (upon advancing) on the liquid front. Following the model of Ref.~\cite{giacomello2015wetting} for nanoscale defects (both chemical and topographical) in the dilute regime, the energy dissipated by defects during an hysteresis cycle is estimated from their wetting free-energy weighted by the area fraction they occupy, see Supplementary Information for details:

\begin{align}
    \text{CAH} &\equiv (\cos \theta_r- \cos\theta_a) \approx (\theta_a-\theta_r) \sin \theta_e\nonumber \\ 
    &=\frac{1}{2} \frac{A_\mathrm{sub}}{A}\lvert(\cos\theta_\mathrm{PDMS}-\cos\theta_\mathrm{sub})\rvert + \frac{1}{2}\frac{A_\mathrm{PDMS}-(A-A_\mathrm{sub})}{A} \lvert \cos\theta_\mathrm{PDMS} \rvert
    \label{eq:CAH}
\end{align}

where $\theta_r$ and $\theta_a$ are the receding and advancing contact angles, the equilibrium contact angle is defined as $\theta_e=(\theta_a+\theta_r)/2$, $A_\mathrm{sub}$ is the surface area occupied by chemical defects, $A_\mathrm{PDMS}$ is the total PDMS area accessible to the contacting liquid, and $A$ is the projected area of the substrate. $\theta_\mathrm{sub}$ and  $\theta_\mathrm{PDMS}$ are the values of the Young contact angle on the bare substrate (silica) and on a flat PDMS layer, respectively.
The factor $1/2$ is due to the assumption that the maximum force a defect can exert on the liquid front, and thus the dissipated energy, coincides, on average, with the defect being half covered by the liquid.
The approximation in Eq.~\ref{eq:CAH}, CAH$\approx (\theta_a-\theta_r)\sin \theta_e$, is valid for small CAH \cite{mchale2022}; for hydrophobic equilibrium contact angles, as PDMS for which $\theta_e = 105^\circ$ \cite{cambiaso2023grafting}, the expression further reduces to CAH$\approx\theta_a-\theta_r$.
Eq.~\eqref{eq:CAH} takes into account both the effect of chemical defects, characterised by the wetting contrast ($\cos\theta_\mathrm{PDMS}-\cos\theta_\mathrm{sub}$) with area fraction $A_\mathrm{sub}/A$ (introduced in Fig.~\ref{fig:dens_profiles}d), and of topographical ones, which result in an excess area $A_\mathrm{PDMS}-(A-A_\mathrm{sub})$ compared to the flat silica surface. This excess area is related to $A_\mathrm{PDMS}/A$ shown in Fig.~\ref{fig:dens_profiles}d. Eq.~\ref{eq:CAH} is an approximate model relating defect wetting energy to CAH, distinct from the usual Cassie-like models that are concerned with the equilibrium contact angle.

\begin{figure}[!ht]
    \centering
    \includegraphics[width=\linewidth]{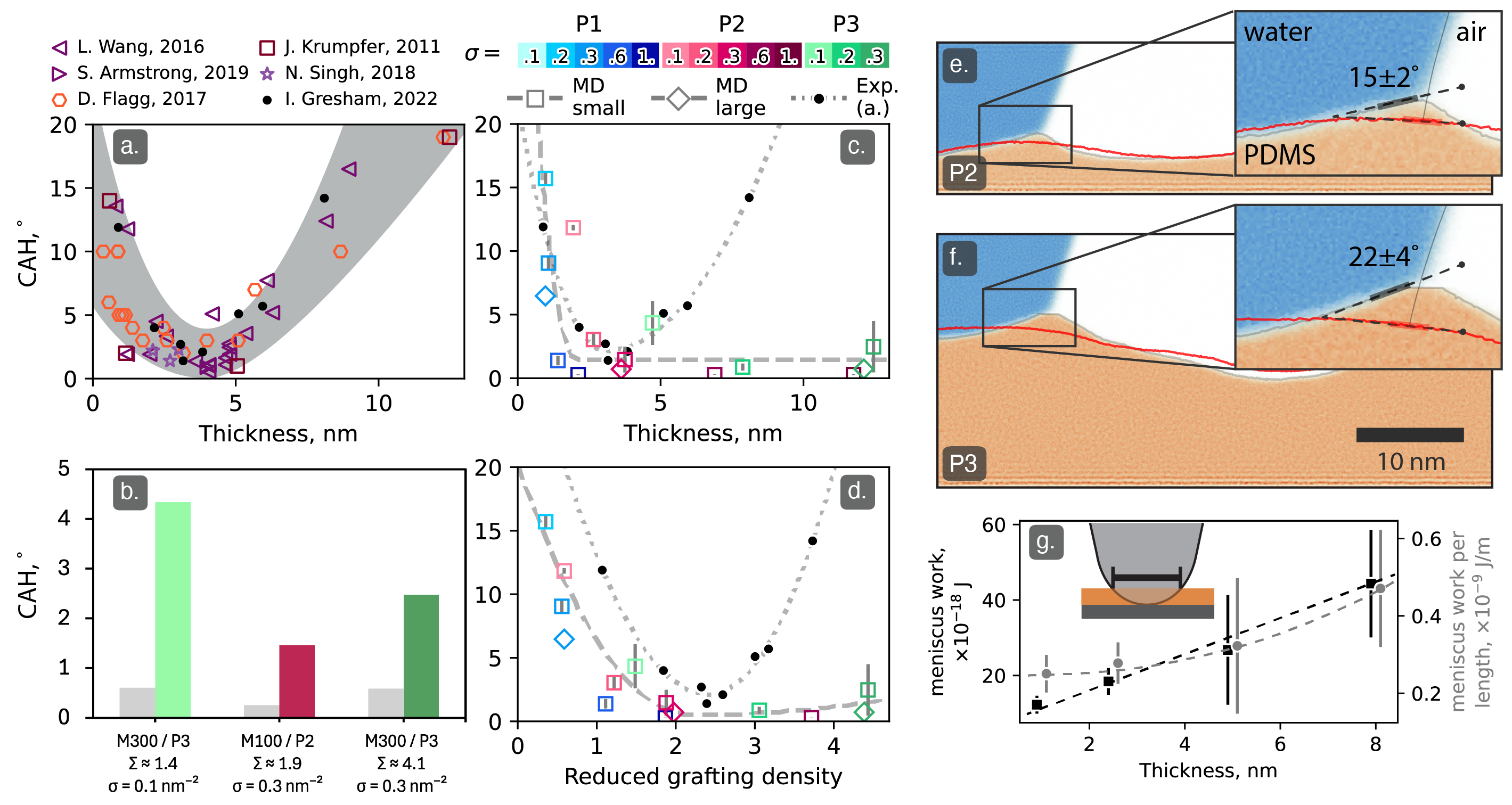}
    \caption{\textbf{Emergence of CAH:}
    a) Contact angle hysteresis as a function of thickness for select literature reports \cite{gresham2023advances}, showing the general trend of CAH vs thickness in CALS.
    b) CAH for a range of polydisperse and monodisperse surfaces as calculated from simulations via Eq.~\ref{eq:CAH}; polydispersity increases the CAH by 1-4 $^\circ$, and as such could explain variation observed in literature for layers of a similar thickness. 
CAH as a function of c) layer thickness and d) reduced grafting density, as obtained from experiments \cite{gresham2023nanostructure} ($\bullet$), and MD simulations ($\square$, $\diamond$) through the use of Eq.~\ref{eq:CAH}.
    Chemical defects explain the general CAH trend for low-$\Sigma$ layers, but the waviness is not sufficient to explain the CAH of high-$\Sigma$ layers.
    Instead, deformation of the PDMS layer around a droplet is likely responsible for the steep increase in CAH at high $\Sigma$.
    e) Simulations of the three-phase contact line show that the (slippery) P2 layer deforms less than the (f) thick P3 layer around a droplet.
    The local slope of the surface changes by (approximately) 15$^\circ$ and 22$^\circ$ for P2 and P3, respectively, accounting for increased CAH in P3 \cite{extrand1996contact}.
    g) Work done on a retracting AFM tip (here termed the meniscus work, left axis) and normalised meniscus work (divided by the circumference of the tip in contact with the layer, right axis) as a function of the PDMS layer thickness, due to deformation of the liquid meniscus upon retraction of the AFM tip.}
    \label{fig:CAH}
\end{figure}

Figure~\ref{fig:CAH}a shows an overview of literature data and serves as a reminder of the two questions posed in the introduction.
Figure~\ref{fig:CAH}b highlights the important observation that, for any layer, polydispersity increases the CAH by 1-4 $^\circ$. This effect may explain the discrepancy in the results in the literature. Using the current synthetic methods, the molecular weight and polydispersity of grafted PDMS chains increase together \cite{gresham2023nanostructure}, and while layer thickness is a useful proxy for chain length, most published reports do not quantify polydispersity. Therefore the variability in published reports on CALS could be a consequence of poor control of polydispersity. Secondly, as described below, polydisperse chains are likely to be more deformable than monodisperse ones, further increasing their CAH.
    
Figure~\ref{fig:CAH}c,d shows CAH computed from Eq.~\ref{eq:CAH} for the simulated systems (modelled CAH), alongside experimental values, plotted against layer thickness $d$ and reduced grafting density $\Sigma$.
The model predicts the steep decline in CAH at low $\Sigma$ (and thickness) seen in experiments, with near quantitative agreement when $\theta_\mathrm{sub}$ is set to $31\degree$. For both experiments and simulations, the CAH minimum ($\approx 1.5^\circ$) occurs in the same range of parameter values ($d=3-6$ nm or $\Sigma =1.5-3$). In both c) and d), the minimum in CAH corresponds to the P2 surface that was shown to have the lowest incidence of chemical defects and waviness. However, experimental and modelled values deviate at higher $\Sigma$ values, with the model in Eq.~\ref{eq:CAH} predicting a rather small increase in CAH due to waviness.

One possible source of this discrepancy is that Eq.~\ref{eq:CAH} assumes rigid topographical defects, whereas the wavy PDMS layer is deformable.
Deformable surfaces are known to increase CAH as compared to their rigid counterparts  because the `ridge' pulled up by capillary forces around the droplet acts as an additional defect \cite{extrand1996contact,limat2012straight,chen2018static,Xiaoteng2024}.
While the effect of ridge formation on the \textit{dynamic} CAH has been investigated \cite{Lhermerout2016, Lhermerout2019}, the mechanism is not as well established for static contact line friction.
The shear modulus of brushes should increase as $\mu\propto 1/N$ \cite{fujii2010shear,fredrickson1992surface}, so it is expected that $\mu(P3)<\mu(P2)<\mu(P1)$.
Our simulations with an explicit three-phase contact line around a droplet show that the ridge pulled up is larger for P3 than for P2 (Fig.~\ref{fig:CAH}e,f), consistent with prior work on a thin monodisperse brush \cite{Badr2022MPIPmolD}. Further simulations of the equilibrated three-phase contact line were conducted for monodisperse layers, revealing that the ridges on monodisperse surfaces exhibit greater symmetry and reduced angularity (Fig.~S5). This indicates that a contact line likely experiences less effective pinning in monodisperse layers and, consequently, reduced CAH.
Experimentally, the work done by the PDMS meniscus on a retracting AFM tip (hence called meniscus work,  measured by integrating the force vs. separation profile \cite{Xiaoteng2024}), was used as a proxy for the energy required to deform the layer at the three-phase contact line.
The average meniscus work measured in our AFM experiments increases with layer thickness, even when correcting for the effect of contact-line length (Fig.~\ref{fig:CAH}g).
If the effect of layer deformation on CAH was exclusively geometric, as postulated in Ref.~\cite{extrand1996contact} following Gibbs pinning criterion \cite{oliver1977resistance}, the additional CAH due to deformation would be significantly higher for P3 than for P2, as the edge angle is $25\degree$ and $13\degree$, respectively (Fig.~\ref{fig:CAH}e,f). Therefore, a significant increase in CAH is expected as the thickness and thus the deformation of the PDMS layer increases: 
this feature could thus account for the increased CAH seen in experiments beyond the Goldilocks zone and will be investigated quantitatively in an upcoming work.

Overall, multiple phenomena explain the Goldilocks zone defining SCALS, as summarised in Fig.~\ref{fig:scals}d.
Chemical defects cause the high CAH generally seen at low thicknesses;
$\Sigma$ is the appropriate parameter to distinguish the chemically heterogeneous patchy regime from the brush one \cite{gresham2023nanostructure}, i.e., the emergence of chemical defects. 
On the other hand, layer deformation accounts for the increase in CAH observed at higher thickness; this mechanism has very recently been suggested by others \cite{Xiaoteng2024, Hauer2024WettingSilicone}.
For a given $\Sigma$, the deformation (and hence CAH) of the layer should correlate with $N$ \cite{fujii2010shear}.
Finally, polydispersity was seen to increase the predicted CAH via an increase in layer roughness across CALS of all thicknesses. Hence, polydispersity could reasonably account for the range of CAH reported in literature for ostensibly similar layers.
Moreover, the flower-like coils induced by polydispersity consist only of long chains, and hence will be more deformable than the rest of the layer.
MD also predicts that layers in the limit of high $\Sigma$ and low polydispersity should also exhibit low hysteresis, regardless of thickness.
These layers are most similar to conventional (tightly packed) self-assembled monolayers. To date, they have not been prepared experimentally.

The explanation of static CAH in terms of the nanoscale features of the grafted surface accounts for some of the important static properties of SCALS, including their water repellency and low droplet adhesion. However, dynamical properties of SCALS, including lubrication and fouling reduction \cite{gresham2023advances}, are not discussed here.
Such properties may be related to the mobility of the layer \cite{gresham2023nanostructure}, which cannot be probed directly in this static model. 
For example, the diffusion of a probe particle could be significantly affected by chemically or topographically heterogeneous surfaces, as reported by fluorescence correlation spectroscopy measurements \cite{gresham2023nanostructure,Xiaoteng2024}.
The anomalous diffusion, however, could be another indirect effect of layer homogeneity rather being the cause of the optimal range of $\Sigma$.

\section{\label{sec:conclusions}Conclusions}

In this work, a multiscale approach combining simulations, experiments, and wetting models was used to connect the molecular composition of grafted PDMS layers to their nanoscale surface topography and to their macroscopic properties, in particular, to contact angle hysteresis.
The mechanism underpinning the Goldilocks zone of SCALS was clarified, i.e., the occurrence of SCALS behaviour in an intermediate range of layer thickness ($3-6$~nm) and reduced grafting density ($\Sigma\approx2$).
On thinner grafted polymer layers, increased pinning of droplet contact line is due to chemical defects, corresponding to exposed patches of the hydrophilic substrate, as the short chain molecules incompletely cover the substrate.
On thicker layers, increased pinning occurs due layer deformation.
For all uniform layers, polymer dispersity leads to the emergence of previously overlooked nanoscale surface waviness, which we confirm here by both experiments and simulations.
This surface waviness can explain the range of CAH reported in literature for layers of similar thicknesses. 
These results, summarised in Fig.~\ref{fig:scals}d, provide design principles for SCALS: 
the layer must uniformly coat the substrate (`just right' $\Sigma$), minimise deformation (`just right' layer thickness) and have low polydispersity. 
Conducting MD simulations informed by experimental parameters leads to sophisticated insights that would be impossible with a single approach in isolation.

\section{\label{sec:methods}Methods}

\subsection{Simulations}
Coarse grained simulations were conducted using GROMACS 2023.3 \cite{abraham2015gromacs} and interactions were modeled using the most recent MARTINI force field (MARTINI 3~\cite{souza2021MARTINI}). The PDMS chain model is described in Cambiaso et al.~\cite{cambiaso2022development}. The simulated systems consist of a hydrophilic substrate ($\theta_Y\sim75^\circ$), representing the chemical state of a silica surface with an intermediate number of silanol groups \cite{kostakis2006effect,cambiaso2023grafting}. The PDMS chains are bonded at random sites, and only one chain can occupy each site. Initially, the chains are in stretched upright positions, then relaxed by steepest descent energy minimization.
For each system, a simulation is carried out in vacuum, using an NVT ensemble, until the PDMS layer reaches a stable average thickness and density profile.
For the cases in which the chains are very long, this process can take a simulation time in the order of $\mu$s.
A synoptic view of the thickness of the simulated PDMS layers is reported in Fig.~S2. Substrates of $50 \times 10$~$nm^2$ were simulated at different grafting density and producing three realization for each chain length distribution, while larger $80 \times 80$~$nm^2$ MD boxes allowed only for a single representative simulation for P1, P2, and P3.
After equilibration in vacuum, the system is solvated and equilibrated again for at least $50$~ns ($100$~ns for the biggest systems), with a mobile piston added to impose a pressure of 1 atm.
The water beads are composed of a mixture of 3 bead sizes, a procedure to avoid artificial ordering of water beads near other surfaces and freezing at room temperature \cite{iannetti2024surface}. 

The length distributions of the polymers in the simulated layers were modelled on experimental distributions obtained by Atomic Force Microscopy (AFM) in Single Molecule Force Microscopy (SMFM) mode \cite{gresham2023nanostructure}. As customary, the polydispersity of polymers is quantified through the Schulz-Zimm distribution:
\begin{equation}
            f(N) = \frac{k^k N^{k-1} e^{-kN} }{ \Gamma(k) }
\end{equation}
where $N$ is the chain length, $\Gamma(k)$ is the gamma function of $k$, and $k$ is a parameter related to the polydispersity index (PDI) through PDI $=1 + 1/k$. 
We sampled the Schulz-Zimm distribution using 3 sets of PDI and average chain length $\overline{N}$ to simulate samples P1, P2, and P3. The PDI and $\overline{N}$ are directly obtained from a fit of the Schulz-Zimm \cite{schulz1939kinetik,zimm1948apparatus} function to the experimental distribution. The actual length distributions after sampling is reported in Fig.~S1. The procedure for calculating surface areas is explained in details in Supplementary information and Fig.~S6

\subsection{Experiments}
The PDMS brushes imaged by AFM were prepared via the approach of Krumpfer and McCarthy \cite{Krumpfer2011RediscoveringSiliconesUnreactive}, following the method documented in our recent work \cite{gresham2023nanostructure}.
Briefly, a few droplets of pure linear PDMS (silicone oil) of different viscosities (P1: 20 cSt, P2: 50~cSt / 350~cSt, and P3: 10 kcSt) were pipetted onto SEMI prime grade silicon wafers (0.675~mm thick, N-type, resistivity 1-20~$\Omega$, orientation $<100>$).
PDMS was allowed to spread across the wafer, before the wafers were placed into an oven and stored at 100~\textdegree C for 24~hours.
PDMS layers were washed with copious amounts of toluene, ethanol, and water, before being soaked in toluene for at least an hour to ensure complete removal of untethered polymer.

AFM measurements (Bruker Multimode 8 AFM with a Nanoscope 5 controller were conducted with Bruker FMV-A tips, which were measured to have a spring constant of approximately 3~N/m.
The Peakforce imaging mode was used to record a 256 by 256 array of force curves over a square area of side either 500 or 1000~nm (pixel length of 2 or 4~nm).
The Peakforce setpoint was 5~nN, Peakforce frequency 2000~Hz,  and scan rate 0.8~Hz.
The Peakforce amplitude, which controls the length of the force curve, was adjusted between $30$ and $60$~nm for each sample, to minimise ringing and ensure an adequate baseline was captured before the tip encountered the surface.
Generally, thicker layers required a higher Peakforce amplitude.
All measurements were conducted on a clean sample in air.

Peakforce mapping produced 65,536 force curves, which were processed in a custom Python script which identified points 1-6 depicted in Fig.~S7-S11 to produce maps of a range of parameters.
Briefly, Peakforce files from the Multimode 8 were imported using the Bruker \textit{Nanoscope} Python module.
Jump-in points were found by interpolating the separation value where the approach curve reached 50\% of the maximum adhesion force value.
Maximum adhesion was taken as the 95\% force percentile in the retraction curve.
Further description, alongside the code required for the analysis, is supplied in Supplementary Information.
The current version of this code is maintained on a Github repository.

\paragraph{Author contributions}
CN procured funding. CN and AG designed research. FR performed simulations. AT and AG supervised simulation work. IJG performed experiments. CN supervised experiments. IJG, FR, AT, and AG analysed data.   
IJG and AG wrote the original draft of the paper; all authors edited the final version.

\begin{acknowledgments}
We thank Glen McHale and Erica Wanless for their insightful comments on the manuscript.
The authors acknowledge funding from the Australian Research Council (FT180100214, DP230100555) and under the National Recovery and Resilience Plan (NRRP), Mission 4, Component 2, Investment 1.1, Call for tender No. 104 published on 2.2.2022 by the Italian Ministry of University and Research (MUR), funded by the European Union – NextGenerationEU– Project Title SoftNanoPores – CUP D53D23002240006 - Grant Assignment Decree No. 957 adopted on 30/06/2023 by the Italian Ministry of Ministry of University and Research (MUR). We acknowledge the EuroHPC Joint Undertaking for awarding access to the EuroHPC supercomputer LUMI, hosted by CSC (Finland) and the LUMI consortium through a EuroHPC Regular Access call.
\end{acknowledgments}

\bibliography{biblio}

\end{document}


\dummylabel{fig:scals}{1}
\dummylabel{fig:dens_profiles}{2}
\dummylabel{fig:CAH}{3}
\dummylabel{fig:chain_lengths}{4}
\dummylabel{tab:Params}{1}

\preprint{SCALS paper}

\title{Supplementary Information for \texorpdfstring{\\}{}
\emph{Molecular origin of slippery behaviour in tethered liquid layers}.}

\author{Fabio Rasera}
\author{Isaac Gresham}
\author{Antonio Tinti}
\author{Chiara Neto}
\author{Alberto Giacomello}

\date{\today}
\maketitle
\section{Simulations}

\subsection{Simulations details}
CG-MD simulations of polydisperse PDMS layers were conducted by modelling interactions with the most recent MARTINI force field (MARTINI 3~\cite{souza2021MARTINI}) in which, on average, two to four heavy atoms and associated hydrogens are mapped into one CG bead. This framework enables the computation of systems larger than those accessible to all-atom models and for longer times, with an overall speedup of at least two orders of magnitude~\cite{dejong2016MARTINI}, while maintaining sufficient chemical and spatial resolution. Given its underlying building block principle, the MARTINI force field has been widely used for simulating polymeric systems\cite{alessandri2021martini,rossi2011coarse}, including grafted polymers\cite{cambiaso2023grafting,rossi2012molecular}.
This approach allowed to simulate for long times sufficiently large surface areas grafted with chemically realistic PDMS \cite{cambiaso2022development} molecules, while scanning a number of relevant constructive and environmental parameters.  The simulations were carried at grafting densities $\sigma$=0.1, 0.2, 0.3, 0.6, and 1.0 gps/nm$^2$. Systematic simulations were performed for smaller domains of size 10~nm by 50~nm, using three replicas at each grafting density, to take into account different arrangements of the PDMS layers. Large individual P1, P2, and P3 surfaces of size 80~nm by 80~nm were also simulated.

 \begin{figure}[h]
     \centering
     \includegraphics[width=1\linewidth]{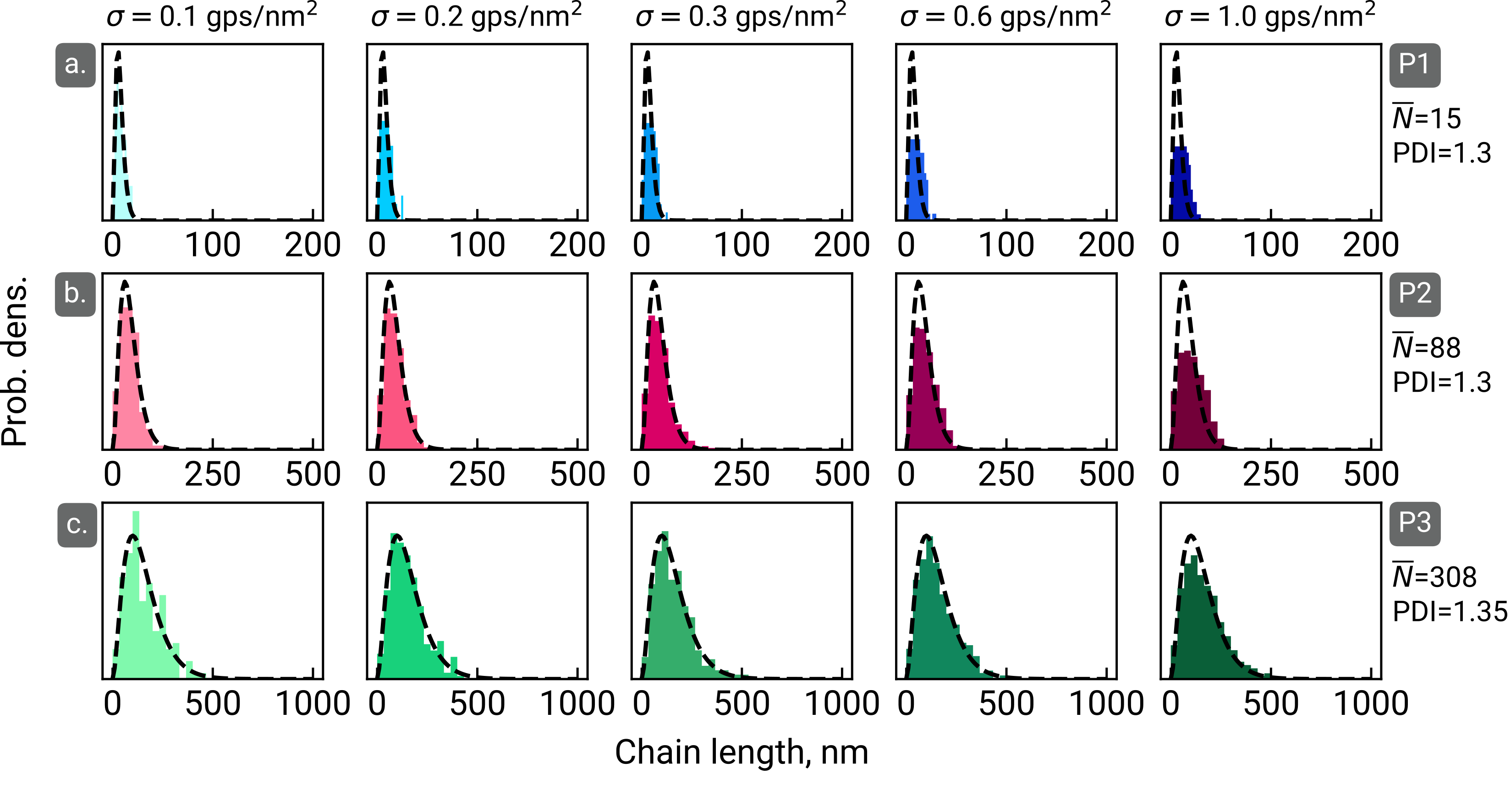}
     \caption{ \textbf{Chain length distributions.} The length distribution of simulated chains (histogram) compared to the Schulz-Zimm distribution obtained from experiments in \cite{gresham2023nanostructure} (line) for a)  P1, b) P2 and c) P3 samples, in the 0.1-1.0 gps/nm$^2$ grafting density range. }
     \label{fig:chain_length_dist}
 \end{figure}

The average layer thickness $d$ was determined by dividing the systems into a 2D grid with a resolution of 1 nm$^2$. The highest atom position within each grid cell was identified, and the arithmetic average of these positions was calculated. Results for both polydisperse and monodisperse layers are presented in Fig.~S2.
 
 \begin{figure}[h]
     \centering
     \includegraphics[width=.8\linewidth]{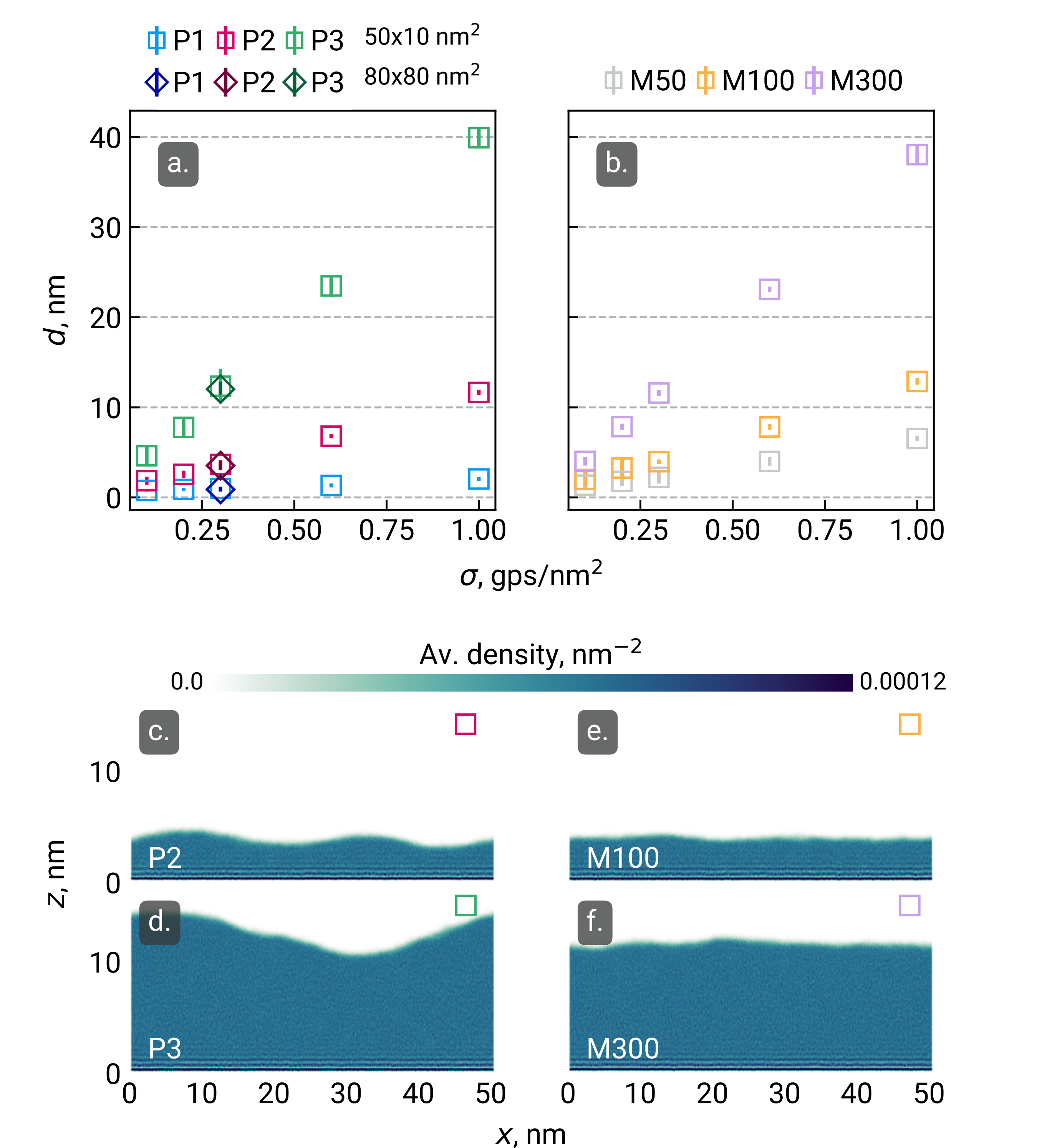}
     \caption{ \textbf{Layer Thickness}: average layer thickness $d$ of simulated PDMS layers. Mean layer thickness of PDMS chains with a) polydisperse and b) monodisperse chain length distribution. The average layer thickness is primarily influenced by the mean length of the composing chains. At an intermediate grafting density (0.3 gps/nm$^2$), the PDMS density during 1 ns of a MD trajectory reveals that (c) P2 and (d) P3 surfaces exhibit a wavy morphology, while (e) M100 and (f) M300 surfaces are relatively flatter, despite having similar average thickness. The P2 surface shows shorter wavelengths and smaller bump heights compared to P3, resembling a monodisperse layer.}
     \label{fig:mean_height}
 \end{figure}

Volume fractions along a given coordinate are valuable for visualizing the packing of a grafted layer and the shape of its interface with a solvent. As shown in Fig.~S3, the volume fractions of PDMS layers in water become more defined as the grafting density increases. Monodisperse layers exhibit a box-shaped volume fraction, indicating a flat interface with water, while polydisperse layers display curved volume fractions, reflecting their irregular surfaces. This contrast is particularly evident when comparing M300 and P3, as P3 features a wavy surface.

 \begin{figure}[!h]
    \centering
    \includegraphics[width=.9\linewidth]{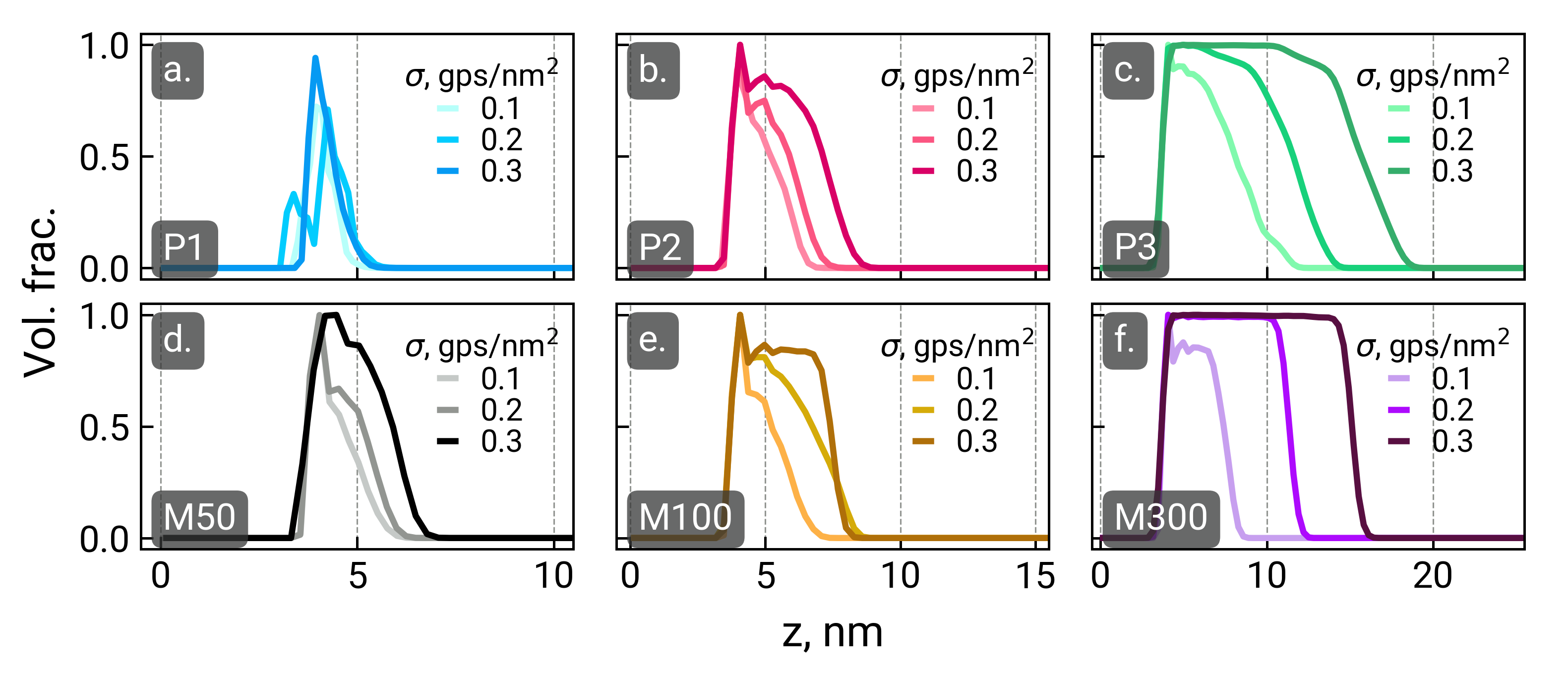}
    \caption{ \textbf{PDMS volume fraction from simulations}: volume fractions of polydisperse (P1, P2, P3) and monodisperse layers (M50, M100, M300) in water were analyzed over a grafting density range of 0.1-0.3 gps/nm$^2$. For the polydisperse cases, volume fractions were averaged across three different realizations, while in the monodisperse cases, a single system was simulated for each chain length and grafting density. In monodisperse samples, the volume fraction assumes a box-like shape once there is sufficient material to form a fully covered surface. Conversely, polydisperse samples exhibit a gradual decrease in volume fraction, reflecting the irregularity of their surface.}
    \label{fig:vol-fracs}
\end{figure}

\clearpage
\subsection{Vertical phase separation}
\begin{figure}
    \centering
    \includegraphics{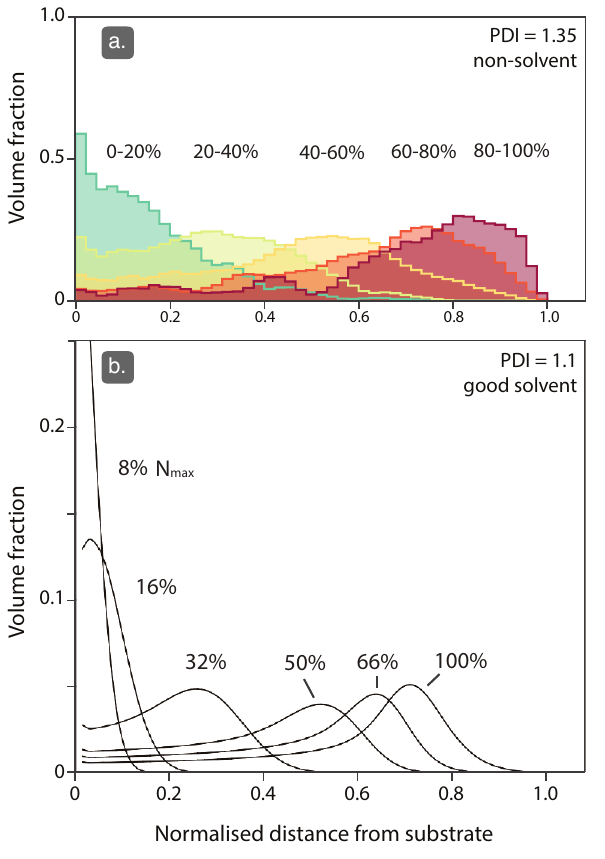}
    \caption{\textbf{Vertical phase separation}: vertical segregation of polymer chains by length, as predicted by the (a) MARTINI coarse-grained molecular-dynamics used here (Figure~3), and (b) the numerical self-consistent field theory of de Vos and Leermakers \cite{devos2009modeling}.
    Both approaches predict that longer chains segregate to the layer periphery, and shorter chains concentrate at the base of the layer.
    The distributions given by de Vos are for a swollen layer (i.e., polymer in a good solvent --- $\chi=0$), while those from the current work are for a collapsed layer (i.e., in poor solvent).
    }
    \label{fig:deVosComparison}
\end{figure}

As discussed in the main paper, the vertical segregation of chains by length in this work (Figure~3) is very similar to that produced by the numerical self-consistent field theory of de Vos and Leermakers \cite{devos2009modeling}.
For ease of comparison, the distributions of different chain fractions from our work and that of de Vos are plotted together in Figure~\ref{fig:deVosComparison}.
The profiles of de Vos are for a polydisperse brush in a good solvent.

\clearpage
\subsection*{Reduced grafting density}

The reduced grafting density $\Sigma$, as defined in Eq.~1 of the main text, can be written for unsolvated polymers as (see Gresham et al.\cite{gresham2023nanostructure} for the derivation):
\begin{equation}\tag{S.1}
    \Sigma = \sigma k \left( \frac{\overline{N}M_m}{\rho}\right)^{2/3} \ ,
    \label{eq:Sigma}
\end{equation}
where $k$ is a constant equal to $\pi \left( \frac{3}{4\pi N_A}\right) ^{2/3}$ and $\sigma$, $\overline{N}$, $M_m$ and $\rho$ are the grafting density, average monomer number of the chains, molecular mass and density. For simulated samples, $\rho$ was calculated using the double cubic lattice method\cite{eisenhaber1995double}, as implemented in the SASA tool from GROMACS.

\clearpage
\subsection*{Substrate deformation}
\begin{figure}[h]
    \centering
    \includegraphics[width=\linewidth]{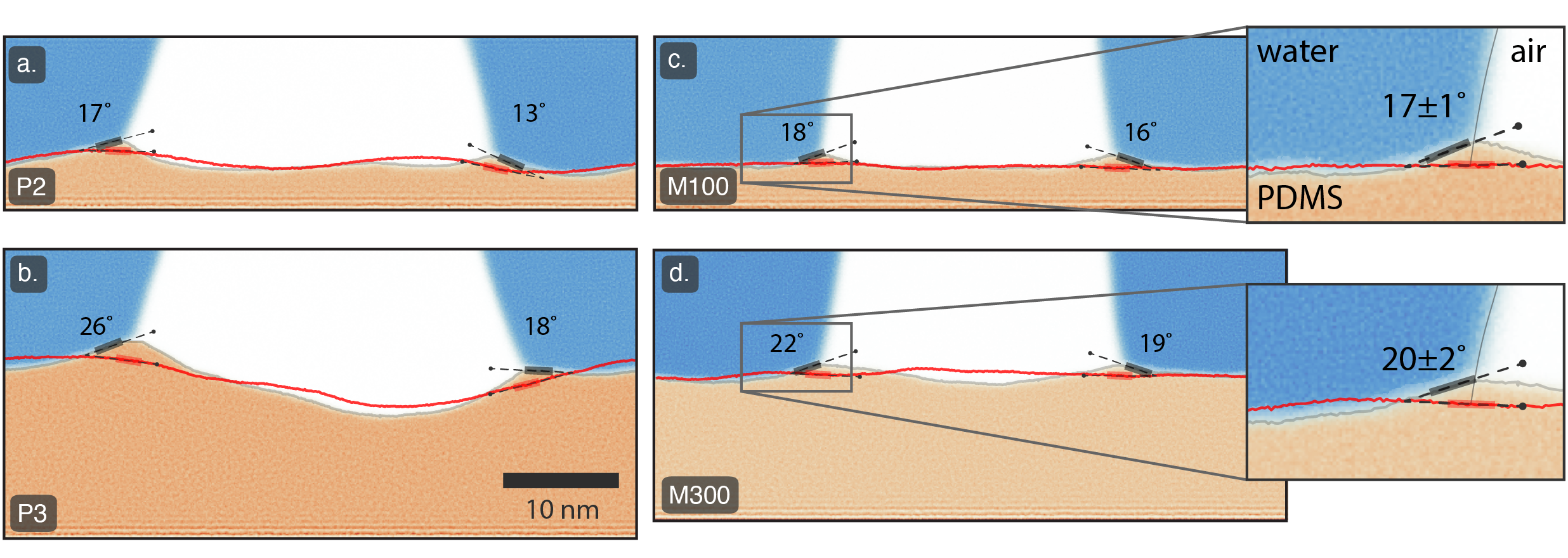}
    \caption{Deformation of the PDMS layer at the three-phase contact line for (a) P2, (b) P3, (c) M100 and (d) M300 substrates.
    Insets are provided for monodisperse layers for direct comparison with Fig.~4; as in Fig.~4 average values are reported in the inset.}
    \label{fig:MonoDefor}
\end{figure}

Fig.~\ref{fig:MonoDefor} shows the deformation of two polydisperse (P2 and P3) and two monodisperse (M100 and M300) substrates, quantified by the angle formed between the local orientation of the surface before and after equilibration with water. The polydisperse P3 exhibit the most pronounced ridge formation, resulting in a significant change of local curvature in comparison with other samples. This is indicative of stronger contact line pinning and greater surface deformation, which correlates with higher contact angle hysteresis (CAH) \cite{extrand1996contact}. In contrast, the monodisperse surfaces (M100 and M300), as well as the polydisperse P2, display a ridge formation that is more symmetrical, resulting in a smaller deformation angle, suggesting that these surfaces experience weaker pinning and, possibly, lower energy dissipation in comparison with P3.

\subsection{\texorpdfstring{$A_{\mathrm{sub}}$}{Asub} and \texorpdfstring{$A_{\mathrm{PDMS}}$}{APDMS} calculation}

\begin{figure}[!h]
    \centering
    \includegraphics[width=.6
\linewidth]{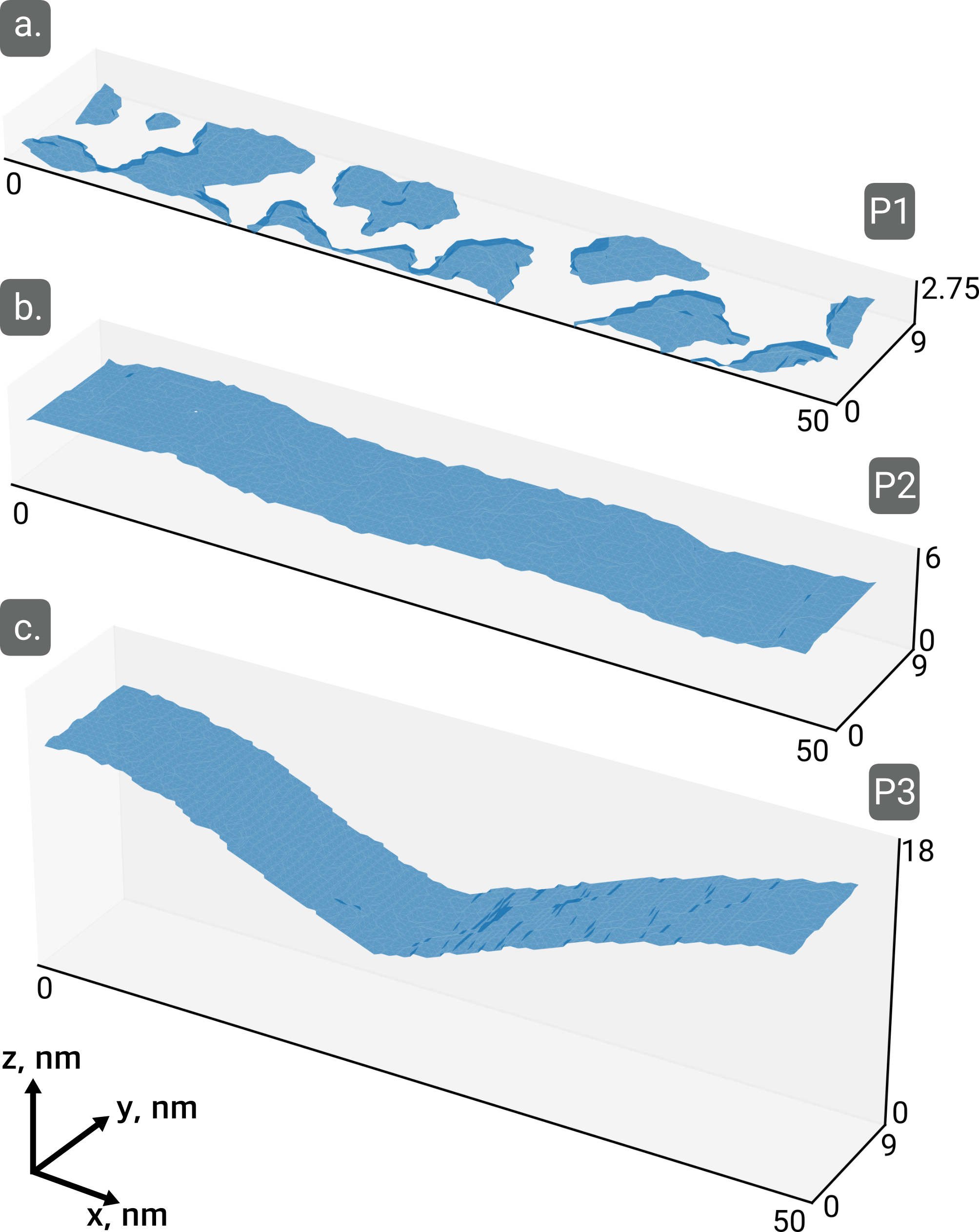}
    \caption{\textbf{Reconstruction of the PDMS surface}: Surfaces reconstruction using marching cubes algorithm. a) P1 realization, rich in chemical defects. b) P2 realization, with approximately flat surface. c) P3 realization, with accentuated waviness.}
    \label{fig:surfaces}
\end{figure}

The marching cubes algorithm\cite{lorensen1998marching}, as implemented in the scikit image library\cite{van2014scikit}, was used to calculate the surface area of the PDMS layers $A_{\mathrm{PDMS}}$. It works by dividing the atom coordinates into a grid of cubes and determining the intersection of a desired isosurface with each cube, creating triangles that approximate the surface topography. $A_{\mathrm{sub}}$ is then calculated as the base area $A$ minus the projected area of the PDMS surface onto the plane parallel to the substrate surface. Figure~\ref{fig:surfaces} shows 3 surfaces reconstructed by coloring the many triangles that compose the surface, where the sum of all triangles' areas is $A_{\mathrm{PDMS}}$.

\clearpage
\section{Modelling}
\subsection{Hysteresis induced by nanodefects}
In order to connect the characteristics of nanoscale defects and macroscopic contact angle hysteresis (CAH), we build upon the framework developed in Ref.~\cite{giacomello2015wetting}. There, the free-energy profile connected with the advancement or retreat of a liquid front across nanoscale defects of chemical or topographical nature was computed. Classical density functional simulations showed that, at such small sizes, the free-energy profile is typically sigmoidal, accounting for either advancing or receding defects, depending on their nature (e.g., chemical defects more hydrophobic than the substrate or more hydrophilic, respectively). Such nanoscale defects can withstand some external force coming from an external pressure or from the unbalanced Young force $\gamma(\cos\theta_\mathrm{app}-\cos\theta_Y)$ of a droplet having an apparent contact angle $\theta_\mathrm{app}$ different from the Young contact angle $\theta_Y$ of the undefected surface. The maximum (minimum) value of such defect force is dictated by the inflection point in the free-energy profile for the advancing/receding process; after this spinodal point is reached, the liquid front jumps beyond the defect and the deformations of the triple line are relaxed \cite{giacomello2015wetting}. 

While capturing the detailed shape of the liquid front and accounting for subnanoscale defects require microscopic calculations, an estimate of the free-energy jump $\delta \Omega^\mathrm{def}=\Omega_\mathrm{wet}-\Omega_\mathrm{dry}$ for overcoming an individual defect can be supplied by macroscopic capillarity. For a chemical defect this reads:
\begin{equation*}
\delta \Omega^\mathrm{chem} = \gamma (\cos\theta_Y-\cos\theta_\mathrm{def}) A_\mathrm{def} 
\end{equation*}
where $\gamma$ is the liquid-vapor surface tension, $\theta_Y$ and $\theta_\mathrm{def}$ the Young contact angles characterizing the undefected surface and the defect, respectively, and $A_\mathrm{def}$ is the area of the defect. Similarly, for a topographical defect, one has:
\begin{equation*}
\delta \Omega^\mathrm{top} = - \gamma \cos\theta_Y A_\mathrm{exc}
\end{equation*}
where $A_\mathrm{exc}$ is the excess area of the defect associated to topographic roughness, i.e., the additional area as compared to the undefected surface. The previous expressions are used in the main text, assuming that the undefected surface coincides with a perfectly flat PDMS layer ($\theta_Y=\theta_\mathrm{PDMS}$)  and that chemical defects have the wetting characteristics of the silica substrate ($\theta_\mathrm{def}= \theta_\mathrm{sub}=75^\circ$). 

The second step in the CAH theory was recognizing that it is possible to estimate the force exerted by a dilute random distribution of defects on a macroscopic triple line, e.g., that of a droplet, from the characteristics of individual defects \cite{Joanny1984}. In order to do this, one integrates the forces coming individual defects at all relative distances between the defect and the unperturbed triple line, taking into account only the stable positions of the triple line, i.e., before the spinodal jumps~\cite{giacomello2015wetting}. This leads to the following expression for the unbalanced Young force acting on the macroscopic triple line: 
\begin{equation*}
    \gamma(\cos\theta_\mathrm{app}-\cos\theta_Y) = - n (\Omega_\mathrm{max} - \Omega_\mathrm{min})
\end{equation*}
where $n$ is the surface density of defects and the subscripts \emph{max} and \emph{min} denote the free energies computed at the maximum and minimum stable position of the triple line, respectively. Using the expression above for the advancing and receding processes, one obtains
\begin{equation*}
   \cos\theta_\mathrm{r}-\cos\theta_a = n \frac{\lvert \delta \Omega^\mathrm{def}_\mathrm{sp} \rvert}{\gamma}
\end{equation*}
where the free energy dissipated when the triple line jumps across each advancing or receding defect, $\lvert \delta \Omega^\mathrm{def}_\mathrm{sp} \rvert$,  is calculated at the spinodal, i.e., at the configuration of the triple line where the maximum or minimum force is attained before the triple line snaps beyond the defect. The expression above allows one to compute CAH given the defect characteristics and their surface density.

The dissipated energy $\lvert \delta \Omega^\mathrm{def}_\mathrm{sp} \rvert$ should be calculated at the inflection point in the free-energy profile. In the main text, since we do not have an explicit calculation of such quantity for each defect, we simply assume that the free-energy profile is a sigmoidal function with height $\delta \Omega^\mathrm{def}$ such that the inflection point is simply achieved at the midpoint and $\lvert \delta \Omega^\mathrm{def}_\mathrm{sp} \rvert \approx \lvert\delta \Omega^\mathrm{def}\rvert /2$. Furthermore, since we only have defects of chemical or topographical nature, the integration over all defects of each kind simply corresponds to summing over the areas in which the substrate is exposed, $A_\mathrm{def}=A_\mathrm{sub}$, for chemical defects and over the excess areas for topographical ones, $A_\mathrm{exc}=A_\mathrm{PDMS}-(A-A_\mathrm{sub})$. The latter expression takes into account the fact that chemical and topographical defects can coexist, in which case the excess area should be computed with respect to only the portion of the projected surface which is covered by  PDMS, $(A-A_\mathrm{sub})$.

\clearpage
\section{Experiments}
\subsection{Imaging CALS with atomic force microscopy}
\begin{figure}[tb]
    \centering
    \includegraphics{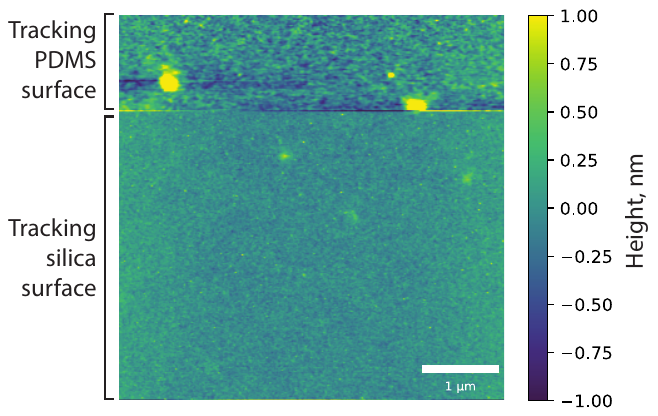}
    \caption{\textbf{Tapping mode AFM}: 5 $\mu$m tapping mode atomic force microscope scan of a 4 nm thick PDMS CALS taken on an Asylum MFP 3D. During the scan, the tip transitioned from tracking the PDMS surface (top) to tracking the silica substrate (bottom).
    When tracking the PDMS surface, the texture appears similar to the wavy surfaces observed in Figure~\ref{fig:dens_profiles}.}
    \label{fig:AFMscan_tapping}
\end{figure}

The meniscus force mapping (MFM) approach described in the work allows for the quantitative and reproducible imaging of tethered liquid surfaces.
The primary reasons why MFM was used instead of conventional tapping-mode atomic force microscopy is that tracking the surface of a thin liquid layer with tapping mode proved to be difficult on the available instrumentation.
In short, the range of `tapping force' (increased either through decreasing the set point or increasing the drive amplitude) required to adequately track the surface of the liquid layer is narrow.
If the tapping force is too low, the tip no longer tracks the surface, if it is too high, the tip instead tracks the silicon substrate.
Consequently, it difficult to track the liquid surface, with the tracked surface often switching mid-image as in Figure~\ref{fig:AFMscan_tapping} without a change in scan parameters.
Further complicating matters, both the silicon substrate and PDMS surface can be exceptionally smooth, which makes differentiating the two surfaces difficult during imaging. 
As such, we found that the MFM technique was much more robust than tapping mode.

Additional reasons MFM was used over tapping mode included the ability to measure both the thickness of the liquid layer and variations in tip-layer adhesion.
Furthermore, the Bruker Peakforce mode allowed for force maps to be acquired at approximately the same rate as conventional tapping mode on the same instrument (approximately 5 minutes an image). 

\clearpage
\subsection{Complete output for AFM measurements}
\begin{figure}[h]
    \centering
    \includegraphics[width=0.95\linewidth]{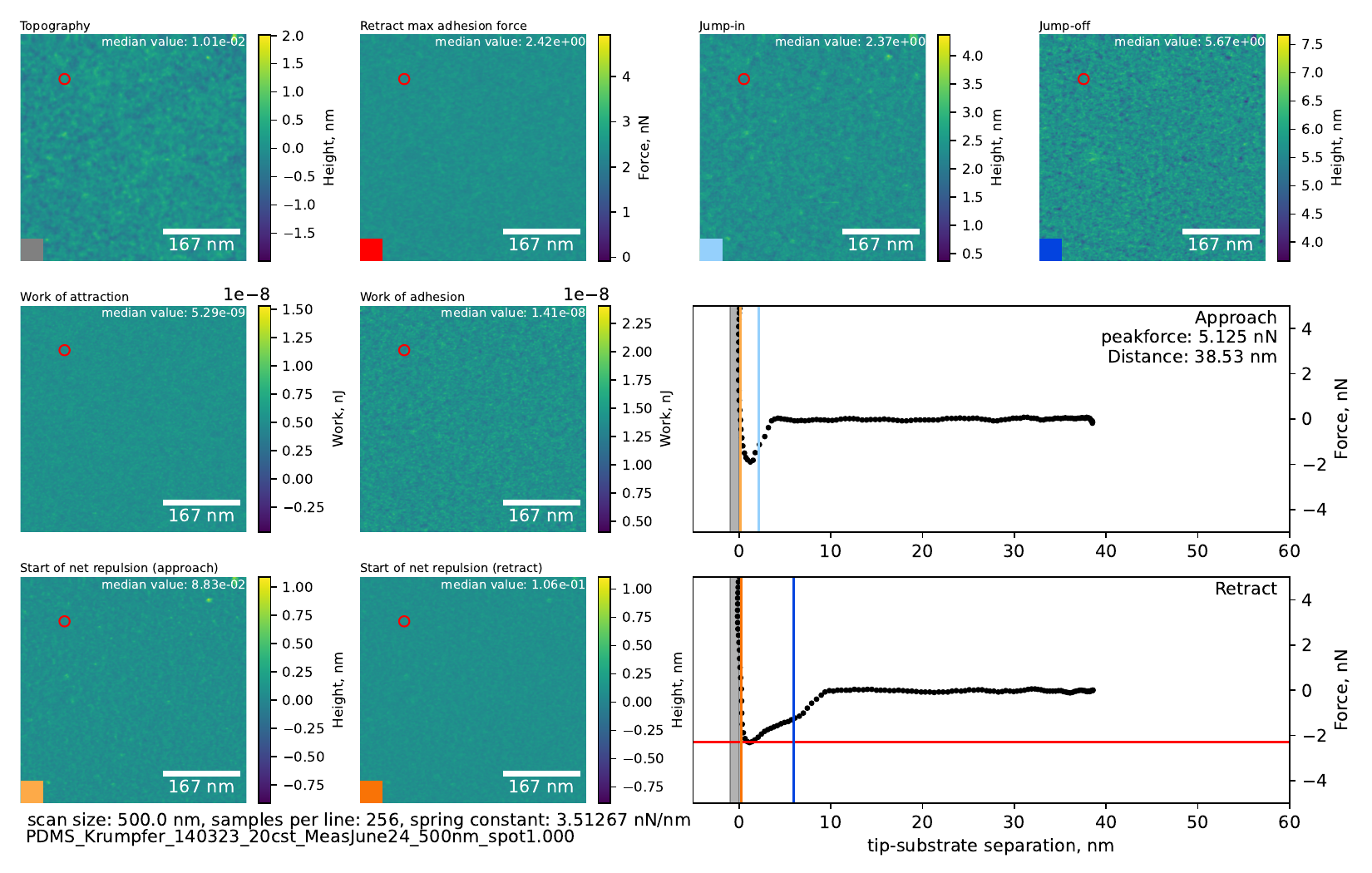}
    \caption{Meniscus force measurements for sample P1 (1 nm thick), 500~nm scan}
    \label{fig:P1500nm}
\end{figure}

\begin{figure}
    \centering
    \includegraphics[width=0.95\linewidth]{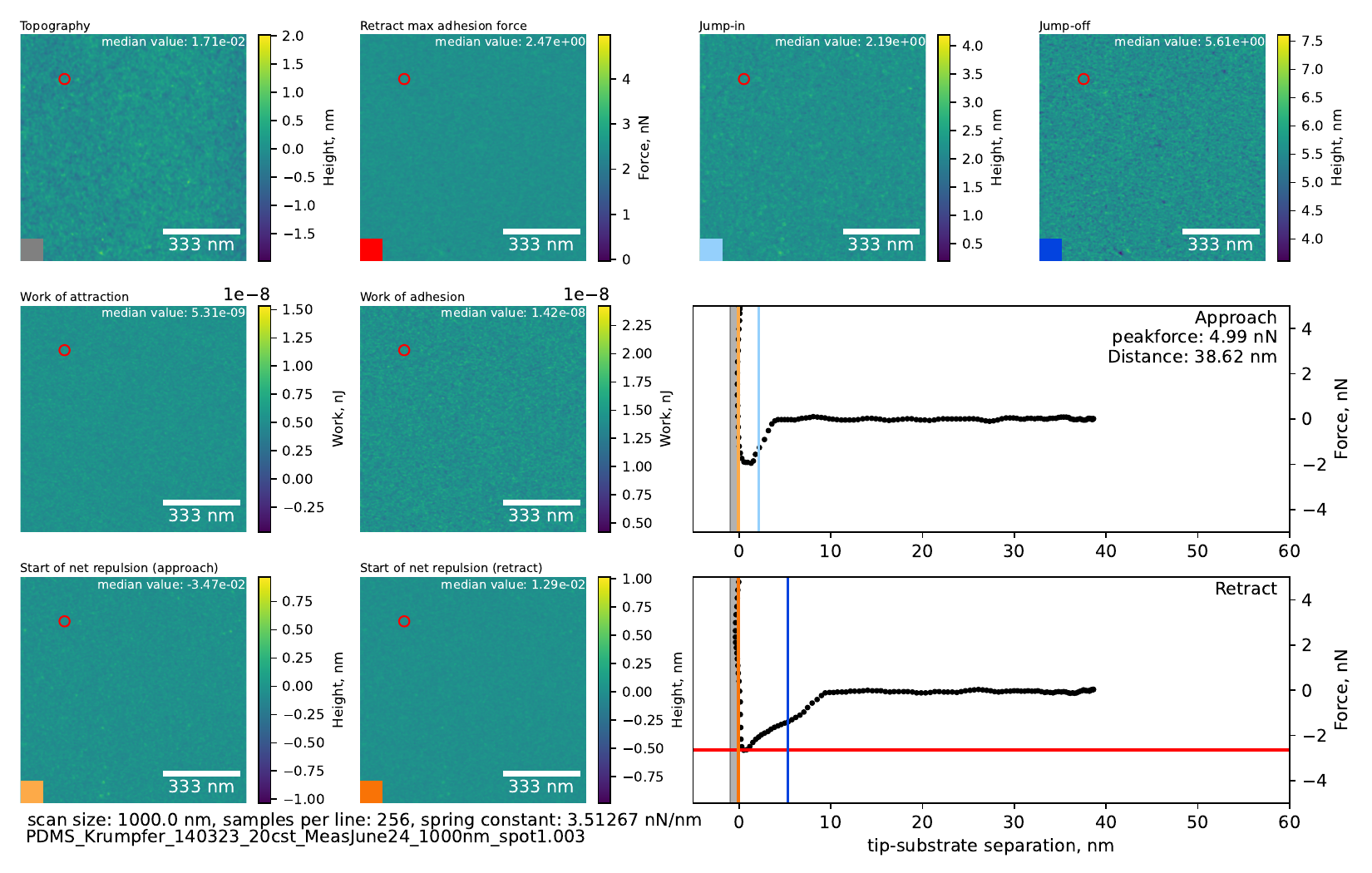}
    \caption{Meniscus force measurements for sample P1 (1 nm thick), 1000~nm scan}
    \label{fig:P11000nm}
\end{figure}

\begin{figure}
    \centering
    \includegraphics[width=0.95\linewidth]{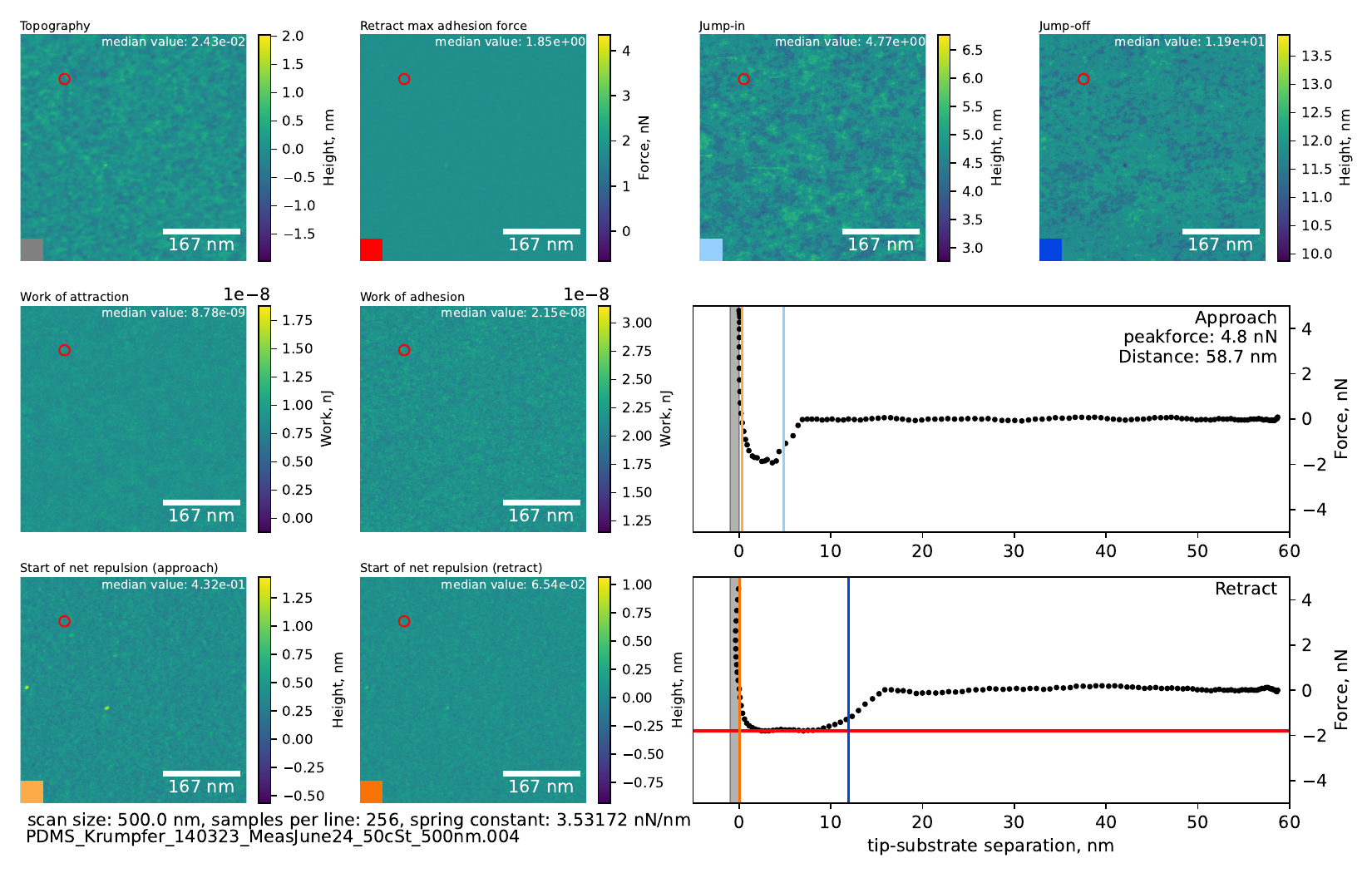}
    \caption{Meniscus force measurements for sample P2 (3 nm thick), 500~nm scan}
    \label{fig:P2a500nm}
\end{figure}

\begin{figure}
    \centering
    \includegraphics[width=0.95\linewidth]{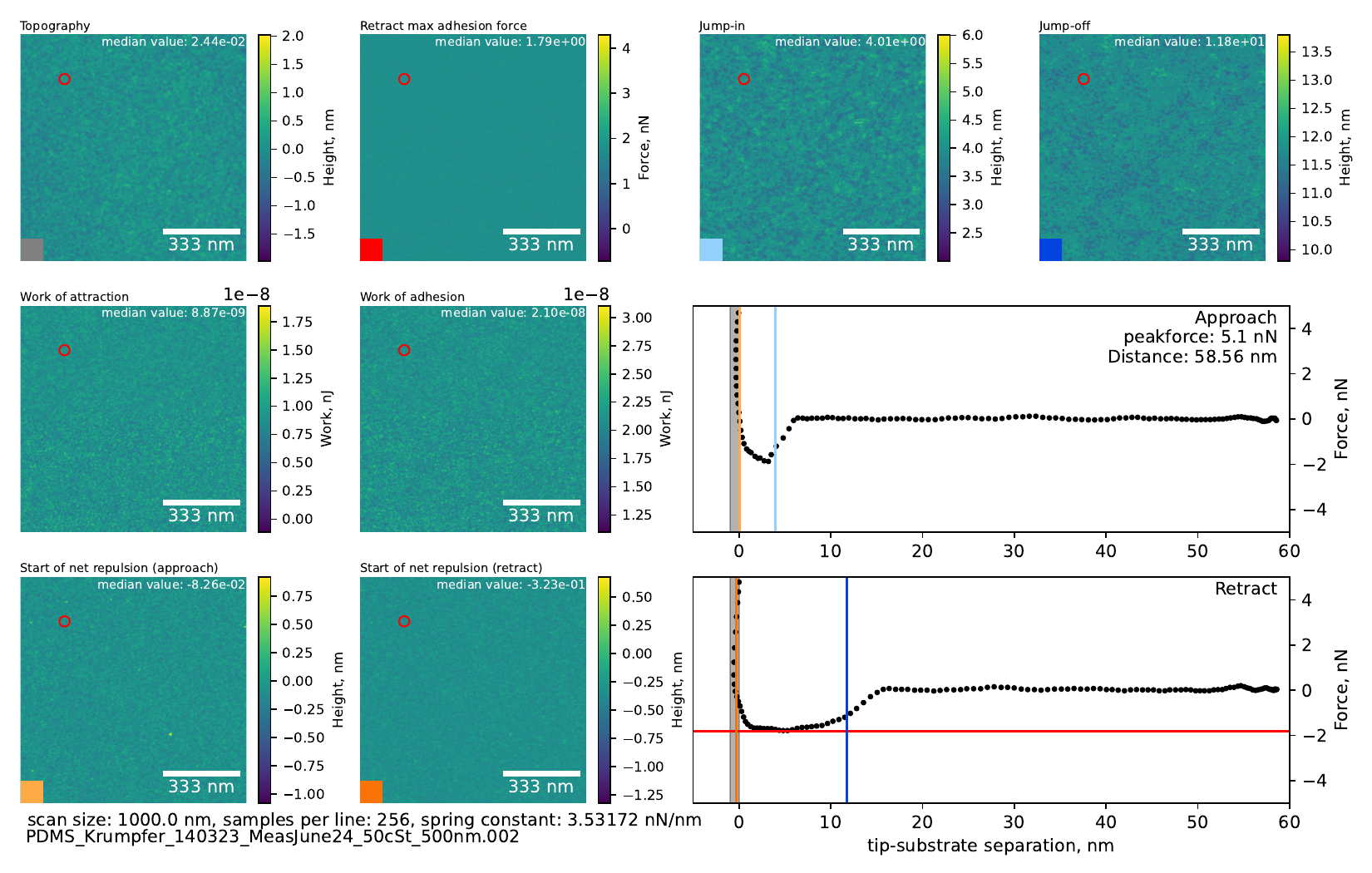}
    \caption{Meniscus force measurements for sample P2 (3 nm), 1000~nm scan}
    \label{fig:P2a1000nm}
\end{figure}

\begin{figure}
    \centering
    \includegraphics[width=0.95\linewidth]{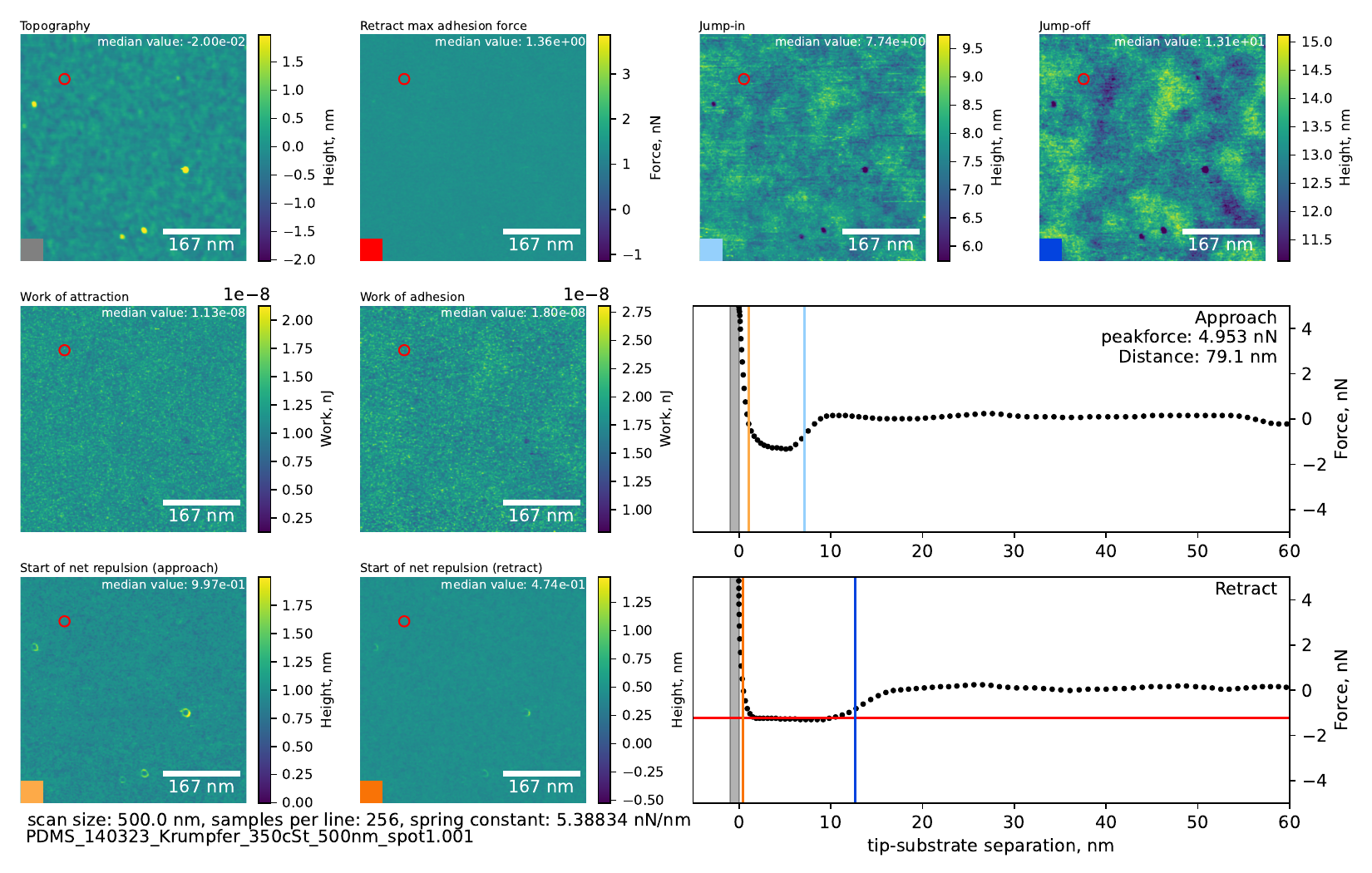}
    \caption{Meniscus force measurements for sample P2 (5 nm thick), 500~nm scan}
    \label{fig:P2b500nm}
\end{figure}

\begin{figure}
    \centering
    \includegraphics[width=0.95\linewidth]{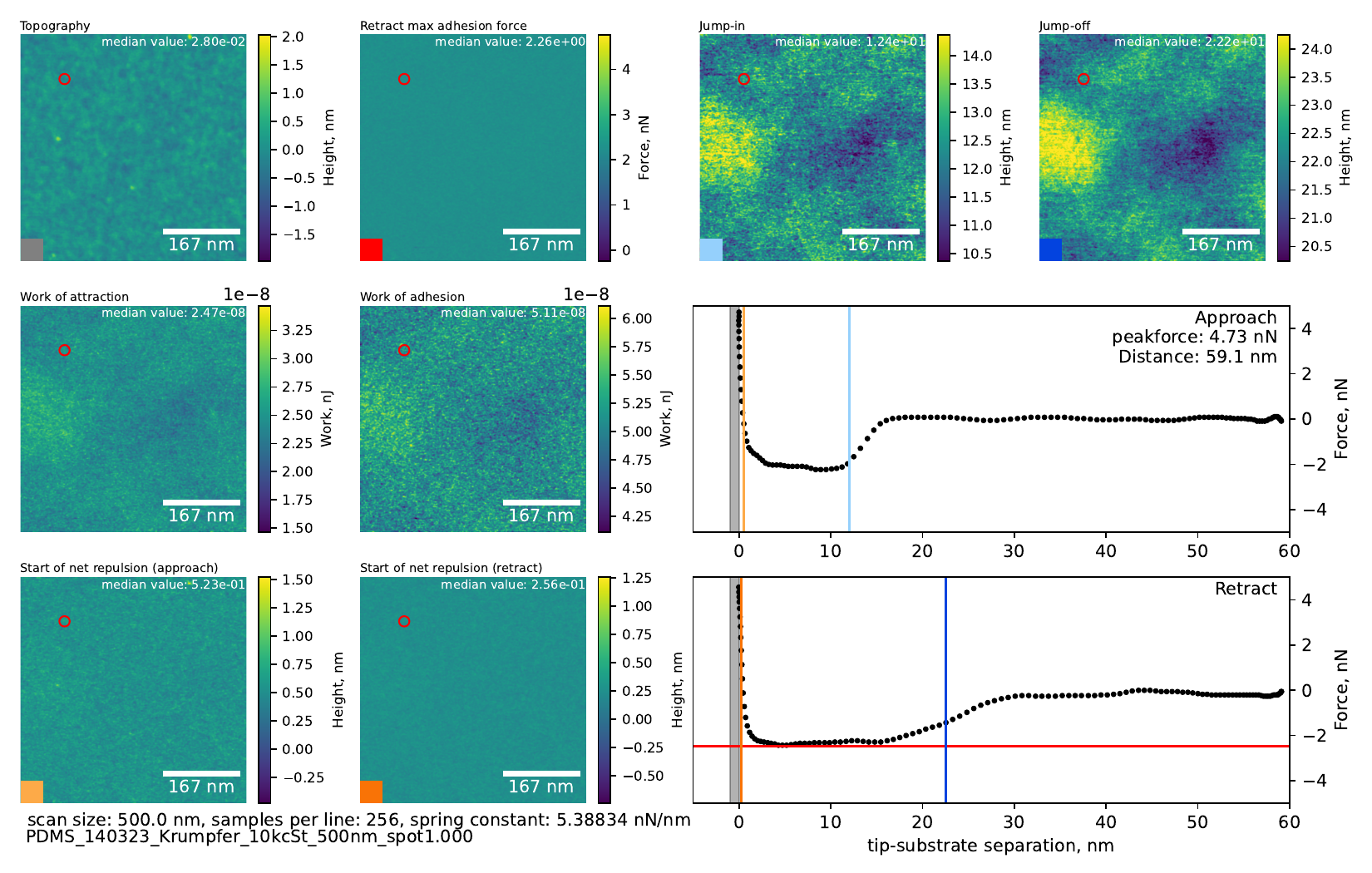}
    \caption{Meniscus force measurements for sample P3 (8 nm thick), 500~nm scan}
    \label{fig:P3500nm}
\end{figure}

\begin{figure}
    \centering
    \includegraphics[width=0.95\linewidth]{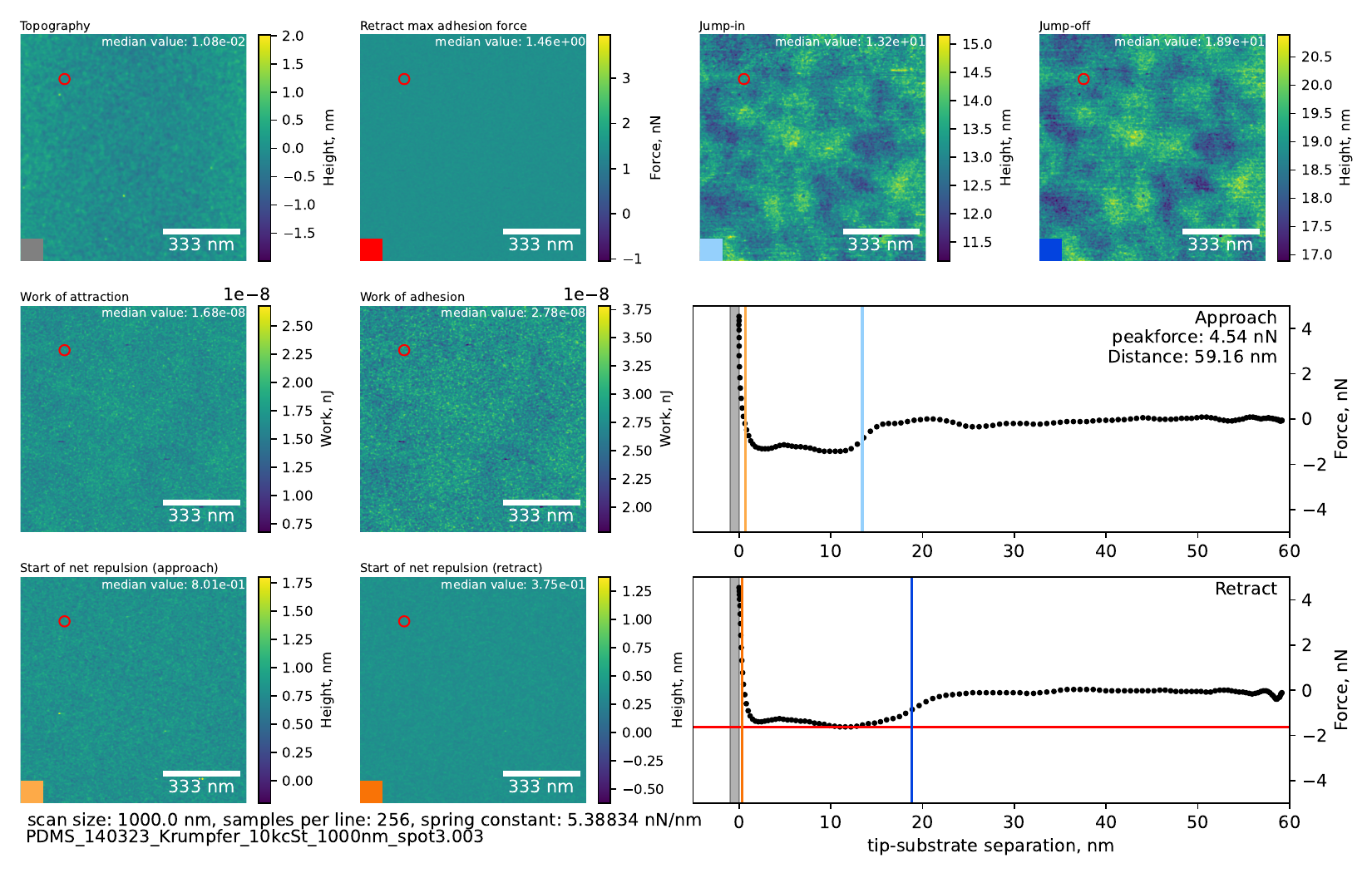}
    \caption{Meniscus force measurements for sample P3 (8 nm thick), 1000~nm scan}
    \label{fig:P31000nm}
\end{figure}

Additional data extracted from meniscus force measurements for the surfaces shown in Figure~\ref{fig:dens_profiles} are presented in Figures~\ref{fig:P1500nm} to~\ref{fig:P31000nm}.
For each surface, 500 and 1000 nm scans are provided.
In each figure, maps of a number of parameters are produced:
\begin{description}
    \itemsep0em
    \item[Topography] The silica substrate topography as defined by the z-position at which the force equals to the peakforce (here 5 nN). The substrate is indicated in the force curves by shaded grey regions.
    \item[Retract max adhesion force] The maximum force exerted on the cantilever upon tip retraction, indicated by the horizontal red line.
    This is used as a measure of chemical heterogeneity in Figure~\ref{fig:dens_profiles}c. These measurements were taken consecutively using the same AFM tip, so a direct comparison of the adhesion values is valid.
    \item[Jump-in] The point at which the cantilever begins to feel meniscus attraction to the sample, indicated by the vertical light-blue line. The jump-in point is taken to correspond to the surface of the tethered-liquid surface. To produce the images in Figure~\ref{fig:dens_profiles}b, the jump-in point is added to the substrate topography.
    \item[Jump-off] The point at which the cantilever ceases to interact with the substrate, indicated by the dark blue line. It has been taken to be proportional to polymer molecular weight in other work \cite{Xiaoteng2024}.
    \item[Work of attraction] The integral of the approach curve between the points of jump-in and net-repulsion.  
    \item[Work of adhesion] The integral of the retract curve between the points of jump-in and net-repulsion.
    The work of adhesion is used in Fig.~\ref{fig:chain_lengths}e as a proxy for the energy required for meniscus formation.
    \item[Start of net repulsion (approach)] separation at which the force acting on the cantilever becomes repulsive (positive) on approach. Values should be close to zero; values far from zero either indicate an elastic substrate or a problem with the measurement (e.g., poorly calibrated optical sensitivity). 
    \item[Start of net repulsion (retract)] separation at which the force acting on the cantilever becomes repulsive (positive) on retract.
\end{description}

\clearpage

\subsection{Code used to process PeakForce data}

The code used to process the PeakForce data into the images presented in Fig.~\ref{fig:dens_profiles}b and Fig.~\ref{fig:P1500nm}-\ref{fig:P31000nm} is presented below.
The code is also hosted on GitHub (\href{https://github.com/igresh/AFMtools}{github.com/igresh/AFMtools}), alongside additional functions to assist in data visualisation.

The code reads a Bruker PeakForce files (with extension \textit{.pfc}) via the \textit{PeakForceImport} function, which produces a numpy binary of the force curve data, a numpy binary of the image provided by the AFM (termed `topography' below) and a \textit{.csv} file of relevant metadata.
These output files are saved in a folder in the output directory; the folder shares the name of the original \textit{.pfc} file.
The force curves provided in the \textit{.pfc} file have already been converted to force vs. separation curves, with zero separation being defined as the point at which the peakforce was reached.
This means that all values calculated from the force curves are relative to the substrate; the substrate topography is also provided in the \textit{.pfc} file.

These initial output files are then processed by the \textit{processForceMap} function, which calculates the parameters below.
Where colours are given in parentheses, they correspond to the lines in the force relevant force curves of Fig.~\ref{fig:P1500nm}-\ref{fig:P31000nm}.
\begin{itemize}
    \itemsep0em
    \item Maximum attractive (i.e., negative) force on approach
    \item Maximum attractive (i.e., negative) force on retract (red)
    \item Jump-in point on approach, interpreted as the start of the liquid layer (light blue)
    \item Pull-off point on retract (dark blue)
    \item Point of net repulsion on approach (light orange)
    \item Point of net repulsion on retract (dark orange)
    \item `Work of attraction', via the integral of the force vs. displacement curve between the points of jump-in and net repulsion on approach.
    \item Work of adhesion, via the integral of the force vs. displacement curve between the points of pull-off and net repulsion on retract.
\end{itemize}

\newpage
\textbf{Imports and helper functions}
\begin{lstlisting}[language=Python]
import numpy as np
import os
import csv
import copy
from scipy.signal import savgol_filter

from nanoscope import files
from nanoscope.constants import FORCE, METRIC, VOLTS, PLT_kwargs, RAW

import sys
import ForceCurveFuncs
import ImageFuncs

def save_array(data, name, directory,savecsv=True):
    np.save(f'{directory}/{name}.npy', data)
    if savecsv:
        np.savetxt(f'{directory}/{name}.csv', data, delimiter=',')

def get_bounds(A, B):
    bmin = np.min([np.mean(A) - np.std(A), np.mean(B) - np.std(B)])
    bmax = np.max([np.mean(A) + np.std(A), np.mean(B) + np.std(B)])
    return bmin, bmax
\end{lstlisting} 
\newpage
\textbf{Opening .pfc files}
\begin{lstlisting}[language=Python]
def PeakforceImport(Filename, output_dir='./Output'):
    """
    Uses the nanoscope module (provided by bruker) to open the peakforce data files.

    Creates a new directory in the output_dir folder using the name of the peakforce
    file. The directory contains the following files:
        -   image.npy (numpy binary) the raw image from the 'image channel'.
            A NxN array containing the topography (height) of the substrate, where N
            is the scan size
            
        -   qnmcurves.npy (numpy binary) the force curves for the image.
            A NxNxYx2 array where N is the scan size and Y is the number of datapoints
            in each force curve.

        -   values.csv (csv file) Relevant imaging parameters.
    """
    arr = []

    if not os.path.exists(output_dir):
        print ('Creating output directory...')
        os.mkdir(output_dir)
            
    name = '.'.join(os.path.basename(Filename).split('.')[:-1])
    print (name)

    if not os.path.exists(f'{output_dir}/{name}'):
        os.mkdir(f'{output_dir}/{name}')
    
    with files.PeakforceCaptureFile(Filename) as file:
        image_channel = file.image_channel
        fv_image, ax_properties = image_channel.create_image(METRIC)
    
        spl = image_channel.samples_per_line
        lines = image_channel.number_of_lines
    
        fc_channel = file.force_curves_channel
        fv_pixels = fc_channel.number_of_force_curves
    
        for pix_idx in range(fv_pixels):
            fz_plot_bl, _ = fc_channel.create_force_z_plot(pix_idx, FORCE)
            fs_plot_bl = fc_channel.compute_separation(fz_plot_bl, FORCE)
    
            arr.append([fs_plot_bl.trace.x,
                        fs_plot_bl.trace.y,
                        fs_plot_bl.retrace.x,
                        fs_plot_bl.retrace.y])
    
        values_of_interest = {'image scan size':image_channel.scan_size,
                              'image scan size unit':image_channel.scan_size_unit,
                              'image line number':image_channel.number_of_lines,
                              'samples per line':image_channel.samples_per_line,
                              'spring constant':image_channel.spring_constant,
                              'optical sensitivity?':image_channel.z_scale_in_sw_units}
    
    arr = np.array(arr, dtype=np.float16 )
    
    np.save(f'{output_dir}/{name}/qnmcurves', arr)
    save_array(data=fv_image, name='image', directory=f'{output_dir}/{name}')

    with open(f'{output_dir}/{name}/values.csv', 'w', newline="", encoding='utf-8') as f:  
        writer = csv.DictWriter(f, fieldnames=values_of_interest.keys())
        writer.writeheader()
        writer.writerow(values_of_interest)
\end{lstlisting} 
\newpage
\textbf{Processing force curves}

\begin{lstlisting}[language=Python]
def processForceMap(direc):
    with open(f'{direc}/values.csv', "r", encoding='utf-8') as infile:
        reader = csv.DictReader(infile)

        for row in reader:
            val_dict = row
        
    scan_size = float(val_dict['image scan size'])
    scan_unit = val_dict['image scan size unit']
    
    arr = np.load(f'{direc}/qnmcurves.npy')

    ExtendsForce = copy.deepcopy(arr[:,0:2])
    RetractsForce = copy.deepcopy(arr[:,2:4])

    baseline_av = np.average(ExtendsForce[:,1,:50], axis=1)
    points_per_line = int(np.sqrt(len(arr)))

    ExtendsForce[:,1] = (ExtendsForce[:,1].T - baseline_av.T).T
    RetractsForce[:,1] = (RetractsForce[:,1].T - baseline_av.T).T

    ExtendsForce[:,1] = ExtendsForce[:,1,::-1]
    ExtendsForce[:,0] = ExtendsForce[:,0,::-1]


    sav_params = {'window_length':5, 'polyorder':1}
    ExtendsForce[:,1] = savgol_filter(ExtendsForce[:,1], **sav_params)
    RetractsForce[:,1] = savgol_filter(RetractsForce[:,1], **sav_params)
    
    
    jump_in = []
    pull_off = []
    wadh_in = []
    wadh_off = []
    rep_on  = []
    rep_off = []

    for idx, [EF, RF] in enumerate(zip(ExtendsForce, RetractsForce)):
        # Calculate jump in
        mask = EF[1]<np.min(EF[1,:50])*0.5
        mask[70:] = False
        if np.sum(mask)==0:
            jump_in.append(0)
        else:
            jump_in.append(np.quantile(EF[0][mask],q=0.98))

        # Calculate work of attraction
        mask = EF[0] < jump_in[-1]+10
        newEF = np.copy(EF)
        newEF[1][newEF[1]>0]=0

        if np.sum(mask)==0:
            wadh_in.append(0)
        else:
            wadh = np.trapz(newEF[1], newEF[0])
            if not wadh == np.inf:
                wadh_in.append(wadh)
            else:
                print ('inf')
                wadh_in.append(0)

        
        # Calculate the location of repulsive onset
        mask = np.logical_and(EF[0] < jump_in[-1], EF[1]>0)

        if np.sum(mask)==0:
            rep_on.append(0)
        else:
            idx_at_int = np.argwhere(mask)[-1][0]
            intercept = np.interp(x=0, xp=EF[1,idx_at_int:idx_at_int+2], fp=EF[0,idx_at_int:idx_at_int+2])
            rep_on.append(intercept)


        # calculate the point of pull-off    
        mask = RF[1]<np.min(RF[1,:50])*0.5
        mask[70:] = False
        if np.sum(mask)==0:
            pull_off.append(0)
        else:
            pull_off.append(np.quantile(RF[0][mask],q=0.98))


        # Calculate work of adhesion
        mask = RF[0] < pull_off[-1]+10
        newRF = np.copy(RF)
        newRF[1][newRF[1]>0]=0

        if np.sum(mask)==0:
            wadh_off.append(0)
        else:
            wadh = np.trapz(newRF[1], newRF[0])
            if not wadh == np.inf:
                wadh_off.append(wadh)
            else:
                print ('inf')
                wadh_off.append(0)


        # Calculate the location of repulsive offset
        mask = np.logical_and(RF[0] < pull_off[-1], RF[1]>0)
        if np.sum(mask)==0:
            rep_off.append(0)
        else:
            idx_at_int = np.argwhere(mask)[-1][0]
            intercept = np.interp(x=0, xp=EF[1,idx_at_int:idx_at_int+2], fp=EF[0,idx_at_int:idx_at_int+2])
            rep_off.append(intercept)


    jump_in  =  np.array(jump_in, dtype=np.float64)
    pull_off =  np.array(pull_off, dtype=np.float64)
    wadh_in  = - 1e-9 * np.array(wadh_in, dtype=np.float64) # report value in nJ
    wadh_off = - 1e-9 * np.array(wadh_off, dtype=np.float64) # report value in nJ
    rep_on   =  np.array(rep_on, dtype=np.float64)
    rep_off  =  np.array(rep_off, dtype=np.float64)

    
    ExtendsForce = np.reshape(ExtendsForce,(points_per_line,points_per_line,2,-1))
    RetractsForce = np.reshape(RetractsForce,(points_per_line,points_per_line,2,-1))


    ExtendsAdh = -np.min(ExtendsForce, axis=3)[:,:,1]
    RetractsAdh = -np.min(RetractsForce, axis=3)[:,:,1]
    wadh_in_arr = np.squeeze(np.reshape(wadh_in,(points_per_line,points_per_line,-1)))
    wadh_off_arr = np.squeeze(np.reshape(wadh_off,(points_per_line,points_per_line,-1)))
    jump_in_arr = np.squeeze(np.reshape(jump_in,(points_per_line,points_per_line,-1)))
    pull_off_arr = np.squeeze(np.reshape(pull_off,(points_per_line,points_per_line,-1)))
    rep_on_arr = np.squeeze(np.reshape(rep_on,(points_per_line,points_per_line,-1)))
    rep_off_arr = np.squeeze(np.reshape(rep_off,(points_per_line,points_per_line,-1)))
    
    save_array(data=ExtendsAdh, name='extends_adhesion', directory=direc)
    save_array(data=RetractsAdh, name='retracts_adhesion', directory=direc)
    save_array(data=wadh_in_arr, name='work_of_attraction', directory=direc)
    save_array(data=wadh_off_arr, name='work_of_adhesion', directory=direc)
    save_array(data=jump_in_arr, name='jump_in', directory=direc)
    save_array(data=pull_off_arr, name='jump_off', directory=direc)
    save_array(data=rep_on_arr, name='net_repulsion_in', directory=direc)
    save_array(data=rep_off_arr, name='net_repulsion_off', directory=direc)
    
    save_array(data=ExtendsForce, name='extend_force_curves', directory=direc, savecsv=False)
    save_array(data=RetractsForce, name='retract_force_curves', directory=direc, savecsv=False)
\end{lstlisting} 

\bibliography{biblio}